\documentclass[amsmath,amssymb,aps,prd,11pt,tightenlines,superscriptaddress,nofootinbib,preprintnumbers,notitlepage]{revtex4-1}

\input{header}

\usepackage[letterpaper,dvips,hmargin={1in,1in},vmargin={1in,1in}]{geometry}

\begin{document}

\preprint{PITT-PACC-2113}
\preprint{UCI-TR-2021-15}

\title{
\vspace*{1.0in}
Discovering Dark Matter at the LHC through Its Nuclear Scattering \\in 
Far-Forward Emulsion and Liquid Argon Detectors 
\vspace*{0.4in}
}

\author{Brian~Batell}
\email{batell@pitt.edu}
\affiliation{Pittsburgh Particle Physics, Astrophysics, and Cosmology Center, Department of Physics and Astronomy, University of Pittsburgh, Pittsburgh, PA 15217, USA}

\author{Jonathan L.~Feng}
\email{jlf@uci.edu}
\affiliation{Department of Physics and Astronomy, University of California, Irvine, CA 92697-4575, USA}

\author{Ahmed Ismail}
\email{aismail3@okstate.edu}
\affiliation{Department of Physics, Oklahoma State University, Stillwater, OK, 74078, USA}

\author{Felix Kling}
\email{felixk@slac.stanford.edu}
\affiliation{Theory Group, SLAC National Accelerator Laboratory, Menlo Park, CA 94025, USA}

\author{Roshan Mammen Abraham}
\email{rmammen@okstate.edu}
\affiliation{Department of Physics, Oklahoma State University, Stillwater, OK, 74078, USA}

\author{Sebastian Trojanowski
\vspace*{0.25in}
}
\email{strojanowski@camk.edu.pl}
\affiliation{Astrocent, Nicolaus Copernicus Astronomical Center Polish Academy of Sciences, ul.~Rektorska 4, 00-614, Warsaw, Poland}
\affiliation{National Centre for Nuclear Research, ul.~Pasteura 7, 02-093 Warsaw, Poland
\vspace*{0.4in}
}

\begin{abstract}
The LHC may produce light, weakly-interacting particles that decay to dark matter, creating an intense and highly collimated beam of dark matter particles in the far-forward direction.   We investigate the prospects for detecting this dark matter in two far-forward detectors proposed for a future Forward Physics Facility: FASER$\nu$2, a 10-tonne emulsion detector, and FLArE, a 10- to 100-tonne LArTPC.  We focus here on nuclear scattering, including elastic scattering, resonant pion production, and deep inelastic scattering, and devise cuts that efficiently remove the neutrino-induced background.  In the invisibly-decaying dark photon scenario, DM-nuclear scattering probes new parameter space for dark matter masses $5~\mev \alt m_{\chi} \alt 500~\mev$.  When combined with the DM-electron scattering studied previously, FASER$\nu$2 and FLArE will be able to discover dark matter in a large swath of the cosmologically-favored parameter space with $\mev \alt m_{\chi} \alt \gev$.  
\end{abstract}

\maketitle

\newpage
\tableofcontents
\newpage

\section{Introduction}
\label{sec:intro}

A primary goal of high-energy colliders is to produce dark matter (DM) particles.  If DM is heavy with a mass near the weak scale, its signature is missing transverse energy, which has been studied in detail for decades.  If DM is light, however, such searches are typically ineffective (as are conventional direct detection searches), and alternative search strategies, experiments, and facilities are needed.

In this study, we consider extremely simple models of light DM in which the Standard Model (SM) is supplemented by a dark photon~\cite{Holdom:1985ag} that decays to pairs of DM particles through $A' \to \chi \chi$.  For dark photons with typical loop-suppressed couplings $\varepsilon \sim 10^{-4} - 10^{-3}$ and $m_{A'} , m_{\chi} \sim \mev - \gev$, the DM annihilates through $\chi \chi \to A'^{(*)} \to f \bar{f}$ in the early universe, yielding the correct thermal relic density.  This model is representative of a broad class of hidden sector theories in which the correct amount of DM is produced through thermal freeze-out within the standard cosmology~\cite{Boehm:2003hm,Pospelov:2007mp,Feng:2008ya,Izaguirre:2015yja,Krnjaic:2015mbs,Batell:2017cmf}, just as in the case of weak-scale DM.  In this scenario, however, the DM is light.  As a result, at colliders, the dark photons and DM are dominantly produced along the beampipe in the far-forward region, escape through holes in collider detectors, and evade all conventional collider searches.  

To remove such ``blind spots'' from the Large Hadron Collider (LHC) physics program, a number of experiments are currently planned for the far-forward region.  FASER~\cite{Feng:2017uoz,Ariga:2018zuc,Ariga:2018pin,Ariga:2018uku} has been completely constructed, and FASER$\nu$~\cite{Abreu:2020ddv,Abreu:2019yak, Abreu:2021hol} and SND@LHC~\cite{Ahdida:2750060} are also being prepared to take data when Run 3 of the LHC begins in 2022.  For the High Luminosity-LHC (HL-LHC) era, a Forward Physics Facility (FPF) is under study~\cite{Feng:2020fpf,FPFKickoffMeeting,FPF2Meeting}.  The FPF would house a suite of far-forward experiments, including possibly FASER2~\cite{SnowmassFASER2}, targeting new long-lived particles that decay visibly in the detector; FORMOSA~\cite{Foroughi-Abari:2020qar}, a milli-charged particle detector; FASER$\nu$2~\cite{SnowmassFASERnu2,SnowmassNeutrinoDetectors}, a 10-tonne emulsion detector; SND2, a successor to SND@LHC; and FLArE~\cite{Batell:2021blf}, a proposed liquid argon time projection chamber (LArTPC) with an active volume of 10 tonnes (FLArE-10) to 100 tonnes (FLArE-100).  FASER$\nu$2, SND2, and FLArE will detect millions of TeV-energy neutrinos, providing a wealth of SM measurements, but they also have the potential to search for light DM and other new particles.  

Here we evaluate the prospects for discovering light DM at FASER$\nu$2 and FLArE through DM-nuclear scattering in the HL-LHC era.  This work complements Ref.~\cite{Batell:2021blf}, which focused on the prospects for observing elastic DM-electron interactions in these detectors; Refs.~\cite{Berlin:2018jbm,Jodlowski:2019ycu}, which explored the potential of FASER to probe inelastic DM; Ref.~\cite{Jodlowski:2020vhr},  which studied the scatterings of unstable, but very long-lived, heavy neutral leptons at FASER$\nu$2; and Ref.~\cite{Boyarsky:2021moj}, which investigated leptophobic DM scattering at SND@LHC.\footnote{See also Refs.~\cite{Batell:2009di,deNiverville:2011it,deNiverville:2012ij,Batell:2014yra,Dobrescu:2014ita,Kahn:2014sra,Coloma:2015pih,deNiverville:2015mwa,deNiverville:2016rqh,Aguilar-Arevalo:2017mqx,Aguilar-Arevalo:2018wea,DeRomeri:2019kic,Dutta:2019nbn,Dutta:2020vop} for studies employing a similar DM search technique at proton beam fixed-target experiments.} We assume these experiments are located in a new cavern that is under study for the FPF, which would place the fronts of these detectors approximately 620 m from the ATLAS interaction point (IP), and we consider 14 TeV $pp$ collisions and the expected HL-LHC integrated luminosity of $3~\text{ab}^{-1}$.  Alternative locations for the FPF that are $\sim 150~\m$ closer or farther from the IP do not change the prospects much, provided, of course, that they are large enough to house the detectors we consider.

We begin by defining the light DM models in \secref{modelswithAprime} and specifying the detectors in \secref{detectors}.  We then consider the dominant processes contributing to DM-nuclear scattering, including elastic scattering ($\chi p \to \chi p$), resonant pion production ($\chi N \to \chi N \pi$), and deep inelastic scattering (DIS) ($\chi N \to \chi X$) in Secs.~\ref{sec:elastic}, \ref{sec:pion}, and \ref{sec:dis}, respectively.  For each of these signals, we devise simple kinematic cuts to differentiate the DM signal from the neutrino-induced SM background.

In \secref{combined}, we then combine all of these DM-nuclear probes with the DM-electron signals investigated in Ref.~\cite{Batell:2021blf}. We find that DM-nuclear scattering and DM-electron scattering are quite complementary, with nuclear scattering more powerful for relatively high masses $m_{\chi} \agt 100~\mev$ and electron scattering more sensitive for low masses $m_{\chi} \alt 10~\mev$.  By combining DM-nuclear and DM-electron scattering, FASER$\nu$2 and FLArE can cover the cosmologically-favored parameter space, where the $\chi$ thermal relic density is at or below $\Omega_{\text{DM}}$, for a wide range of DM masses between $\mev \alt m_{\chi} \alt \gev$.  In \secref{combined}, we also compare the sensitivity of FASER$\nu$2 and FLArE to non-LHC experiments that have discovery potential for invisibly decaying dark photons and light DM~\cite{Battaglieri:2017aum,Beacham:2019nyx}.  Our conclusions are presented in \secref{conclusions}.

\section{Invisibly-Decaying Dark Photon Models}
\label{sec:modelswithAprime}

In this section, we describe two popular benchmark models in which light DM interacts with the SM through an invisibly decaying dark photon mediator.  Given its coupling to electrically charged particles and quarks, in particular, the dark photon efficiently mediates scattering between DM and nuclei, making these models an interesting test case for our study. 

The dark photon, $A'$, is a massive gauge boson that arises when the SM is supplemented with a new broken U(1)$_D$ symmetry. For light GeV-scale dark photons, the dark photon Lagrangian is
\be
\mathcal{L} \supset -\frac{1}{4} F'_{\mu\nu}F^{'\mu\nu} +\frac{1}{2} m_{A'}^2 A'_\mu A^{'\mu} + A'_\mu \left(\varepsilon \, e \, J_{EM}^\mu + g_D \, J_D^\mu \right) ,
\ee
where $F'_{\mu\nu}$ is the dark photon's field strength, $m_{A'}$ is the dark photon mass, $\varepsilon$ is the kinetic mixing parameter, $J_{EM}^\mu$ and $J_D^\mu$ are the SM electromagnetic and U(1)$_D$ currents, respectively, and $g_D \equiv \sqrt{4 \pi \alpha_D}$ is the U(1)$_D$ gauge coupling. 

For the DM candidates, $\chi$, we will examine two popular examples: Majorana fermion DM and complex scalar DM. The corresponding Lagrangians are
\be
\mathcal{L} \supset 
\begin{cases}
\displaystyle{\ \frac{1}{2} \overline \chi i \gamma^\mu \partial_\mu \chi 
-\frac{1}{2} m_\chi \overline \chi \chi} \quad \text{(Majorana fermion DM)} \\
\vspace{-10pt}\\
\displaystyle{\ |\partial_\mu \chi|^2 - m_\chi^2 |\chi|^2} \quad \text{(complex scalar DM)} \ , \\
\end{cases}
\ee
where $m_\chi$ is the DM mass. The $U(1)_D$ currents associated with these models are 
\be
\label{eq:JD}
J_D^\mu = 
\begin{cases}
\displaystyle{\ \frac{1}{2} \overline \chi \gamma^\mu \gamma^5 \chi} \quad \text{(Majorana fermion DM)} \\
\vspace{-10pt}\\
\displaystyle{\ i \chi^* \overset{\text{\footnotesize$\leftrightarrow$}}{\partial^\mu} \chi} \quad \text{(complex scalar DM)}  \ .
\end{cases}
\ee

These two DM models have many similarities, but also some key differences.  We discuss them in turn, beginning with the Majorana fermion case.  As noted in \secref{intro}, an attractive feature of these light DM models is the fact that the observed DM relic density can be easily obtained through thermal freeze-out.  For $m_{A'} > 2 m_\chi$, Majorana fermion DM annihilates in the early universe through $\chi \chi \to A^{'(*)} \to f \bar f$ with cross section
\be
\label{eq:sigv}
\sigma v \propto \alpha \, v^2 \, \frac{\varepsilon^2 \, \alpha_D \, m_\chi^2}{m_{A'}^4}  = \alpha \, v^2 \, \frac{y}{m_\chi^2} \ ,
\ee
where we have assumed $m_{A'} \gg m_\chi$ and $y \equiv \varepsilon^2 \alpha_D (m_\chi/m_{A'})^4$~\cite{Izaguirre:2015yja}. As evident from \eqref{sigv}, the annihilation is $P$-wave, and so bounds from cosmic microwave background (CMB) temperature anisotropies on late-time DM annihilation are not very constraining in these models~\cite{Ade:2015xua,Slatyer:2009yq}.  In addition, the scattering of Majorana fermion DM in direct detection experiments is also velocity-suppressed at the non-relativistic energies relevant for these searches, and so direct detection null results also do not set strong limits.  

For complex scalar DM, the annihilation cross section is, in fact, similar to that for Majorana fermion DM. \Eqref{sigv} still applies, and so the complex scalar DM model also evades CMB bounds.  In contrast to the Majorana fermion case, however, the non-relativistic scattering of complex scalar DM in matter is not velocity-suppressed.  Direct detection null results are therefore a significant constraint on this model.  These bounds may be evaded, however, if a small mass splitting is introduced to make the DM scattering transition inelastic~\cite{TuckerSmith:2001hy}.

In this work, we will present our results in the $(m_\chi, y)$ plane.  As we will see, at the relativistic energies relevant for the LHC, the DM-nuclear interactions for Majorana fermion and complex scalar DM are very similar, and so the results we derive will be almost imperceptibly different in the $(m_{\chi}, y)$ plane.  We will therefore simply present the Majorana fermion DM results.  At the same time, to understand the cosmological significance of these results, we will also present ``thermal targets,'' the regions of parameter space where the thermal relic density is identical to the observed DM abundance.  These differ slightly for the Majorana fermion and complex scalar DM models, and so we will present both, using the relic density predictions of Ref.~\cite{Berlin:2018bsc}.  

To reduce the parameter space to two dimensions, we will present results for $\alpha_D = 0.5$ and $m_{A'}  = 3 m_\chi$ throughout this work. These represent relatively conservative choices in terms of characterizing the experimental prospects for testing the thermal freeze-out hypothesis, at least in the regime $m_{A'} \gg m_\chi$.  Of course, if $m_{A'} - 2 m_\chi \ll m_{A'}$, the annihilation rate is resonantly enhanced, and the corresponding thermal targets occur at smaller couplings and can be much more challenging to probe at colliders~\cite{Feng:2017drg,Berlin:2020uwy,Bernreuther:2020koj}.

\section{Detectors and Simulation \label{sec:detectors}}

\subsection{Benchmark Detectors \label{sec:detect}}

The benchmark detectors we consider are identical to those studied in Ref.~\cite{Batell:2021blf}, except that they are now assumed to be housed in the ``new cavern'' FPF, placing them 620 m from the ATLAS IP.  We review their most salient characteristics here; for more details, see Ref.~\cite{Batell:2021blf}.

FASER$\nu$2~\cite{SnowmassFASERnu2} is envisioned to be a larger version of FASER$\nu$~\cite{Abreu:2020ddv}, currently being built for LHC Run 3. The FASER$\nu$2 benchmark detector we consider here is a 10-tonne rectangular tungsten-emulsion detector with location and size given by
\begin{equation}
\text{FASER$\nu$2}: \ L = 620~\m \, ,\ \Delta = 2~\m \,  , \ S_T = (0.5~\m \times 0.5~\m) \, ,
\label{eq:FASERnu2}
\end{equation}
where $L$ is the distance from the IP to the front of the detector, and $\Delta$ and $S_T$ are the longitudinal and transverse dimensions of the tungsten target. At the ATLAS IP during the HL-LHC, it is expected that the beam half-crossing angle will vary by as much as 250 $\mu$rad, moving the beam collision axis horizontally by as much as 15 cm at the detector location. Given the detector's transverse dimensions and the $\sim 20~\cm$ spread of the DM signal and neutrino background~\cite{Kling:2021gos}, the crossing angle will have little effect on our results; for simplicity, we assume that the detector is always centered on the beam collision axis.

We will assume that tracks down to momenta of 300 MeV can be detected and that the emulsion is exchanged periodically so that the track density remains manageable.  This requires changing the detector every $30~\text{fb}^{-1}$ or so, or less if a sweeper magnet is available to bend away muons produced at the IP.   

The main disadvantage of emulsion detectors for this DM search is the lack of timing, which makes it difficult to reject muon-induced backgrounds.  To remedy this, it is necessary to augment the tungsten-emulsion detector with interleaved electronic tracker layers, which would provide event time information.  This design follows the successful example of the OPERA experiment~\cite{Acquafredda:2009zz}, and an analogous design is being implemented for SND@LHC~\cite{Ahdida:2020evc}.  We will, therefore, assume that muon-induced backgrounds can be rejected by vetoing events in coincidence with a high-energy muon track. 
It is important to note, however, that all of our FASER$\nu$2 sensitivities depend on this assumption, and if muon-induced backgrounds are difficult to reject in emulsion detectors, liquid argon technology may be preferable for dark matter detection.

For FLArE, we consider two sizes with physical dimensions
\begin{eqnarray}
&& \text{FLArE-10 (10 tonnes)}: \ \ \quad  L = 620~\m,\ \Delta = 7~\m,\ \ \, S_T = (1~\m \times 1~\m) \ ,
\label{eq:LAr10ton} \\
&& \text{FLArE-100 (100 tonnes)}: \ \ L = 620~\m, \ \Delta = 30~\m,\ S_T = (1.6~\m \times 1.6~\m) \ ,
\label{eq:LAr100ton}
\end{eqnarray}
where, as above, $L$ is the distance from the IP to the front of the detector, $\Delta$ and $S_T$ are the longitudinal and transverse dimensions of the active volume, and we assume that the detector is centered on the beam collision axis. 

Particle kinetic energy thresholds for LArTPC detectors typically lie in the 10-100 MeV range.  For protons, we will consider two kinetic energy thresholds: a conservative value of 50 MeV, as is considered in the DUNE Conceptual Design Report~\cite{Acciarri:2015uup}, and a more optimistic choice of 20 MeV. Concerning the latter, we note that the ArgoNeuT experiment has already achieved thresholds for such short proton tracks down to 21 MeV~\cite{Acciarri:2014gev,Acciarri:2018myr}. For other particles, including shower-like objects (electrons, photons, neutral pions) and charged pions, we will assume a 30 MeV kinetic energy threshold, which is broadly consistent with Refs.~\cite{Acciarri:2014gev,Acciarri:2015uup,Acciarri:2018myr}. In contrast to emulsion detectors, LArTPCs have good active event timing capabilities, particularly when equipped with a light collection system~\cite{Acciarri:2016smi,ICARUS:2020wmd}, and we will assume that vetoing events with a coincident muon is sufficient to remove all muon-induced backgrounds.  

\subsection{Expected Neutrino Fluxes \label{sec:nufluxes}}

A crucial ingredient for the estimation of background rates is the flux of neutrinos passing through the different detectors. We use the dedicated forward physics event generator \texttt{Sibyll~2.3c}~\cite{Ahn:2009wx, Riehn:2015oba, Riehn:2017mfm}, as implemented in the CRMC simulation package~\cite{CRMC}, to simulate the primary collisions. We then use the fast neutrino flux simulation introduced in Ref.~\cite{Kling:2021gos} to simulate the propagation of SM hadrons through the LHC beam pipe and magnets and their decays into neutrinos.

\begin{figure}
    \centering
    \includegraphics[width=0.99\textwidth]{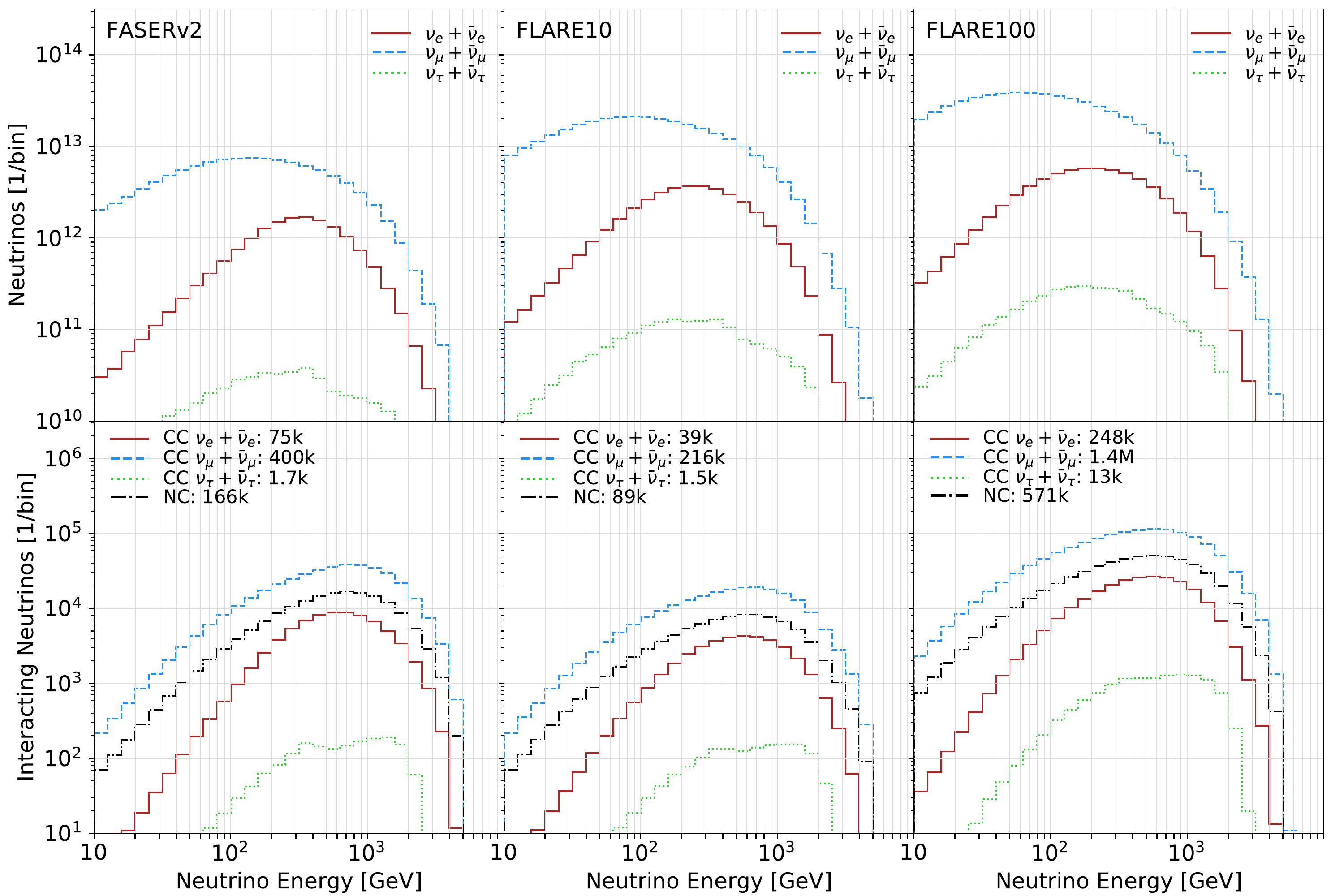}
    \caption{The number of neutrinos passing through the detector (top) and interacting in the detector (bottom), for FASER$\nu$2 (left), FLArE-10 (center), and FLArE-100 (right) during the HL-LHC era. The detector geometries and locations are described in the text. These results assume 14 TeV $pp$ collisions and an integrated luminosity of $\mathcal{L}=3~\iab$ and are estimated using \texttt{Sibyll 2.3d} and the fast neutrino flux simulation introduced in Ref.~\cite{Kling:2021gos}.
    \label{fig:neutrinoflux}}
\end{figure}

The results are presented in \cref{fig:neutrinoflux} for the HL-LHC with an integrated luminosity of $3~\iab$,
assuming no beam crossing angle. The upper panels show the numbers of neutrinos passing through the detectors. Unsurprisingly, detectors with a larger cross sectional area will have more neutrinos passing through them. The lower panels show the numbers of charged current (CC) and neutral current (NC) DIS neutrino interactions in the detectors, where we use the neutrino interaction cross sections from Ref.~\cite{Abreu:2019yak}. Note that the event rate is larger for FASER$\nu$2 than FLArE-10, despite the two detectors having the same mass. This is because the neutrino beam is strongly collimated around the beam collision axis, and so a narrow detector with more mass close to the beam axis, such as FASER$\nu$2, will observe a larger event rate. During the HL-LHC era, we expect about $3.9\times 10^4$ electron neutrino, $2.2\times 10^5$ muon neutrino, and $1.5\times 10^3$ tau neutrino CC interactions in the FLArE-10 detector. In addition, we expect about $8.9\times 10^4$ NC neutrino interactions. The average energy of these interacting neutrinos is about $600~\gev$. 

In addition to the total neutrino interaction rates that, for each flavor, are dominated by DIS, we also provide in \cref{tab:CCQEetalrates} the expected number of events for several exclusive scattering channels. These include both CC quasi-elastic and NC elastic scatterings (denoted in the table by CCQE and NCEL, respectively), as well as the relevant resonant pion production channels (CCRES and NCRES). We estimate them by convoluting the above neutrino fluxes with the cross sections simulated with \texttt{GENIE}~\cite{Andreopoulos:2009rq,Andreopoulos:2015wxa}. As can be seen, in total approximately 3000 CCQE and CCRES and 1000 NCEL and NCRES events are expected in FLArE-10 during the entire HL-LHC era, and the scattering rate is about $30\%$ larger for FASER$\nu$2, and a factor of $7-8$ larger for FLArE-100. These events are mainly due to interactions of the muon neutrinos, while electron neutrinos are responsible for about $10\%$ of the event rates, and tau neutrinos give subdominant contributions. 

\begin{table*}[t]
\setlength{\tabcolsep}{5.2pt}
\centering
\begin{tabular}{c||c|c|c|c|c|c||c|c|c|c|c|c||c|c}
  \hline
  \hline
  &\multicolumn{6}{c||}{CCQE} & \multicolumn{6}{c||}{CCRES} & NCEL & NCRES
  \\ Detector
  & $\nu_e $    
  & $\bar\nu_e$    
  & $\nu_\mu$    
  & $\bar\nu_\mu$    
  & $\nu_\tau$
  & $\bar\nu_\tau$
  & $\nu_e $    
  & $\bar\nu_e$    
  & $\nu_\mu$    
  & $\bar\nu_\mu$    
  & $\nu_\tau$
  & $\bar\nu_\tau$
  & all
  & all
  \\
  \hline
  FASER$\nu$2
  & 57
  & 50
  & 570
  & 355
  & 1.9
  & 1.6
  & 170
  & 183
  & 1.6k
  & 1.1k
  & 5.4
  & 5.1
  & 170
  & 1.3k
  \\
  \hline
  FLArE-10
  & 43
  & 40
  & 425
  & 260
  & 2.0
  & 1.6
  & 120
  & 140
  & 1.2k
  & 860
  & 5.6
  & 5.1
  & 130
  & 940
  \\
  \hline
  FLArE-100
  & 325
  & 290
  & 3.3k
  & 2k
  & 20
  & 15
  & 930
  & 980
  & 9.2k
  & 6.8k
  & 54
  & 50
  & 980
  & 6.5k
  \\
  \hline
  \hline
\end{tabular}
\caption{Expected event rates for charged current quasi-elastic (CCQE), charged current resonant (CCRES), neutral current elastic (NCEL), and neutral current resonant (NCRES) interactions of neutrinos in the FASER$\nu$2, FLArE-10, and FLArE-100 detectors. The results for CC interactions are given for each neutrino flavor separately, while, for the NC events, all the contributions are summed up.}
\label{tab:CCQEetalrates}
\end{table*}

As discussed in Ref.~\cite{Kling:2021gos}, the neutrino fluxes predicted by different commonly-used event generators are somewhat different, indicating a flux uncertainty of about a factor of 2. This situation will improve in the coming years, given dedicated theoretical efforts to reduce these uncertainties; see, e.g., Ref.~\cite{Bai:2020ukz}. In addition, measurements of the energy spectra of CC neutrino interactions at FASER$\nu$ and SND@LHC during LHC Run~3 and later in the FPF neutrino detectors will provide direct measurements of the neutrino fluxes. In the following, we, therefore, assume that the neutrino flux uncertainties are dominated by statistical uncertainties.

\subsection{Signal Modeling\label{sec:modeling}}

Given our chosen benchmark scenario with $m_{A^\prime} = 3\,m_\chi$, the DM particles originate from the decays of on-shell dark photons produced at the ATLAS IP. We simulate the flux of DM particles through the far-forward detectors with the geometries and locations given in Eqs.~(\ref{eq:FASERnu2}), (\ref{eq:LAr10ton}), and (\ref{eq:LAr100ton}), normalizing the number of events to the total integrated luminosity of $\mathcal{L}=3~\iab$ anticipated for the HL-LHC era. The dark photons produced in rare $\pi^0$ and $\eta$ decays are obtained by employing the \texttt{CRMC} simulation package~\cite{CRMC} and the dedicated \texttt{EPOS-LHC} Monte Carlo tool~\cite{Pierog:2013ria}. In addition, we include dark photon production by dark bremsstrahlung, using the Fermi-Weizsacker-Williams approximation, following the discussion in Refs.~\cite{Blumlein:2013cua,deNiverville:2016rqh,Feng:2017uoz}.

A rich variety of DM-nuclei scattering processes can be studied with the far-forward detectors.  To organize the discussion, in the following, we will divide them into distinct categories in a way similar to neutrino interactions; see Ref.~\cite{Formaggio:2013kya} for a review. We first study the case of elastic DM-nucleon scattering, which leads to events with single proton charged tracks in the detector. Next, we consider the exclusive inelastic processes of resonant pion production produced through DM-nucleon interactions. Finally, we consider DM-induced DIS, which is most relevant at high-momentum transfer.

\section{Elastic Scattering \label{sec:elastic}}

\subsection{Signal} 

Here we consider elastic DM-nucleon scattering and the associated signature of a single proton track in the detector with no additional visible charged tracks emerging from the interaction vertex. As mentioned above, we will also assume that there is no through-going muon in the detector that could be associated with the DM-induced event. When presenting the results, we will further require that the proton momentum, $p_p$, be above a minimum value defined by the energy threshold of the detector (see \secref{detectors}) and below a maximum value that we chose to maximize the signal to background ratio, $S/\sqrt{B}$. 

The single proton signature is most directly associated with elastic scatterings of DM off protons, $\chi p \to \chi p$. The relevant differential cross section is~\cite{Leitner,Batell:2014yra}
\begin{equation}
\label{eq:dffxsecDMQE-Majorana}
\frac{d\sigma(\chi p \to \chi p)}{dQ^2} = \frac{4 \pi \varepsilon^2 \alpha \alpha_D Q^2}{(E_\chi^2 - m_\chi^2)(m_{A'}^2+Q^2)^2}\left[ A(Q^2)  + \left(\frac{E_\chi}{Q} - \frac{Q}{4m_N}\right)^2 (\tilde F_{1,p}^2+\tau \tilde F_{2,p}^2) \right],
\end{equation}
where $E_\chi$ is the incoming DM energy, $Q^2 = 2 m_p (E_p-m_p)$ is the squared four-momentum transfer with $E_p$ the outgoing proton energy and $m_p$ the proton mass, and
\begin{equation}
   A(Q^2) = 
   \begin{cases}
 \displaystyle{ \left[ \frac{1}{4} \tilde F_{1,p}^2 \left(1 \! - \! \frac{1}{\tau}\right) +\frac{1}{4}\tilde  F_{2,p}^2  (1\! -\! \tau) +  \tilde{F}_{1,p}\tilde{F}_{2,p} \right] \left(\tau + \frac{m_\chi^2}{m_p^2} \right)}  \ \text{(Majorana fermion DM)}
   \\
\displaystyle{-\frac{1}{4}\,(\tilde{F}_{1,p}+\tilde{F}_{2,p})^2\,\left(\tau+\frac{m_\chi^2}{m_p^2}\right)} \ \text{(complex scalar DM)} \ ,
   \end{cases} 
\label{eq:Afactor}
\end{equation}
with $\tau =  Q^2/(4 m_p^2)$. The proton form factors can be expressed as
\begin{equation}
\tilde F_{1,p}(Q^2) = \frac{ 1 + \mu_p\tau}{1+\tau} G_D(Q^2) \ ,\hspace{1cm}\tilde F_{2,p}(Q^2) = \frac{\mu_p -1 }{1+ \tau } G_D(Q^2) \ ,
\end{equation}
where $\mu_p = 2.793$, and
$G_D(Q^2) = (1+ Q^2/M^2)^{-2}$, with $M = 0.843\, {\rm GeV}$.

As advertised in \secref{modelswithAprime}, the scattering cross sections for Majorana fermion and complex scalar DM have the same high-energy limit. This is evident upon inspection of \eqsref{dffxsecDMQE-Majorana}{Afactor}, which reveals that the first term proportional to $A(Q^2)$ in \eqref{dffxsecDMQE-Majorana} is negligible compared to the second term for large $E_\chi$. The projected exclusion bounds presented below are therefore valid for both the Majorana fermion and complex scalar DM scenarios. We also note that the integrated cross section becomes independent of the DM energy at large $E_\chi$.

Additional signal events could arise from elastic DM scatterings off neutrons, $\chi n\to \chi n$, in which the outgoing neutron re-scatters before leaving the nucleus and produces a final-state proton.  
The relevant cross section for this process can be obtained from \eqsref{dffxsecDMQE-Majorana}{Afactor} by replacing the proton mass and form factors with the quantities appropriate for neutrons~\cite{Batell:2014yra}. However, because the dark photon mediator couples to electric charge, its coupling to neutrons vanishes in the limit of zero momentum transfer. Therefore, for the models considered here, elastic DM-neutron scattering is considerably suppressed relative to elastic DM-proton scattering. Similarly, inelastic DM scattering followed by the absorption of all charged tracks and neutral pions inside the nucleus, besides a single outgoing proton, contributes subdominantly to the total DM signal event rate. We have verified this using \texttt{GENIE}, under the assumption that the impact of nuclear final-state interactions (FSI) on such particles in DM-induced events can be well approximated by their impact on neutrino events with the same momentum transfer to the nucleus. 

In addition to the outgoing proton's energy, its direction can also be observed.  Angular cuts were found in Ref.~\cite{Batell:2021blf} to be useful in separating DM-electron scattering from neutrino-electron scattering, but they are less useful here.  In DM-electron scattering, the additional discriminating power was related to the mass hierarchy between the target electron and the incoming DM particles, $m_e\ll m_\chi$. For the DM-nuclear scattering considered here, however, $m_\chi \alt m_p$ in the parameter space of interest, and so the DM particles behave similarly to essentially massless neutrinos. In the following, we will therefore focus only on the energy cut.  

Elastic scatterings $\chi p\to\chi p$ generally lead to low visible energy depositions due to the strong form factor suppression for large momentum transfers, $Q^2\gtrsim 1~\gev^2$. As a result, we will typically set the maximum outgoing proton momentum, $p_{p}^{\textrm{max}}$, to values below $1~\gev$. The DM detection prospects for this signature improve with softer lower limits on the outgoing proton momentum. For this reason, FLArE can be more sensitive than FASER$\nu$2 if the FLArE proton kinetic energy threshold, $E_{k,p}$, can be lowered to 20 MeV, as discussed in \cref{sec:detect}. Below, we present in detail the estimated sensitivity reach and background estimates for both types of detectors. 

\subsection{Neutrino-Induced Backgrounds}

The dominant neutrino-induced backgrounds to DM-nucleon elastic scattering come from neutral current elastic scatterings (NCEL) of all three neutrino flavors that produce the outgoing proton in the final state, $\nu p \to \nu p$. Additional background events can be induced by deep inelastic neutrino scatterings (NCDIS) and resonant pion production processes (NCRES), in which, occasionally, most of the outgoing particles are absorbed in the nucleus due to FSI. We assume below that outgoing electrons and muons can be sufficiently discriminated from protons so that CC neutrino interactions can be neglected in the background discussion.

In \cref{tab:elasticbackgrounds}, we present the total background event rates obtained with \texttt{GENIE} for FASER$\nu$2, FLArE-10, and FLArE-100. In the case of liquid argon detectors, we impose a selection cut on the minimum proton kinetic energy of either $E_{k,p}>20$ or $50~\mev$, corresponding to the assumed proton detection thresholds discussed in \cref{sec:detect}. The latter condition corresponds to a minimum proton momentum of $p_p\gtrsim 300~\mev$, which we also require in the analysis for the emulsion detector. We also cut on the maximum proton momentum, $p_p < p_p^{\rm max} = 1~\gev$, and for the more optimistic proton threshold in FLArE, $E_{k,p}>20~\mev$, we additionally study a more aggressive upper momentum cut, $p_p^{\textrm{max}} = 500~\mev$. Finally, in each case, we veto on events containing any additional charged tracks or neutral pions emerging from the nucleus, besides the single proton, that have energies above their corresponding detection thresholds; see \cref{sec:detectors}. As can be seen, in the HL-LHC era, the expected number of background events can be roughly 100 events for FLArE-10 and 1000 events for FLArE-100. 

\begin{table}
\begin{center}
\begin{tabular}{ c | c||c||c } 
\hline
\hline
\multicolumn{2}{c||}{\textbf{Elastic} $\chi p\to \chi p$} & \multicolumn{1}{c||}{$\nu$-induced backgrounds} & \multicolumn{1}{c}{DM: $m_\chi=100~\mev$, $\varepsilon=6\times 10^{-4}$}\\
\hline
FASER$\nu$2 & $p_p>300~\mev$, $p_p<1~\gev$ & $310$ & $34$\\
\hline
         & $E_{k,p}> 20~\mev$, $p_p<500~\mev$ & $100$ & $37$\\ 
FLArE-10 & $E_{k,p}> 20~\mev$, $p_p<1~\gev$ & $125$ & $42$\\ 
         & $E_{k,p}> 50~\mev$, $p_p<1~\gev$ & $120$& $23$\\ 
\hline
          & $E_{k,p}> 20~\mev$, $p_p<500~\mev$ & $810$ & $260$\\ 
FLArE-100 & $E_{k,p}> 20~\mev$, $p_p<1~\gev$ & $1050$ & $310$\\ 
          & $E_{k,p}> 50~\mev$, $p_p<1~\gev$ & $1010$ & $165$\\ 
\hline
\hline
\end{tabular}
\end{center}
\caption{Neutrino-induced background and DM signal events for the single proton signature for several choices of selection cuts on the outgoing proton momentum $p_p$. We assume 14 TeV $pp$ collisions with integrated luminosity $3~\iab$. The cuts on the minimum proton momentum are dictated by the assumed experimental thresholds, as discussed in \cref{sec:detect}. The maximum proton momentum is set to $1~\gev$ for FASER$\nu$2. For FLArE-10 and FLArE-100, we also consider an additional case with $p_p<500~\mev$. The DM signal corresponds to the benchmark scenario with parameters $(m_\chi, \varepsilon) = (100~\mev,6\times 10^{-4})$, $m_\chi=m_{A^\prime}/3$, and $\alpha_D=0.5$, and takes into account the efficiency factors (see text).
}
\label{tab:elasticbackgrounds}
\end{table}

The number of background events in FASER$\nu$2 is between those in the two liquid argon detectors. The surprisingly large number of expected background events in FASER$\nu$2 when compared with FLArE-10, which has a similar mass, is mainly driven by the additional impact of neutrino-induced NCRES events that mimic the single proton signal. The outgoing pions produced in these events often have energies corresponding to the mass difference between the dominant $\Delta$ resonance and the proton, $E_{\pi} \sim m_{\Delta} – m_p \sim 300~\mev$. As a result, such events often lead to pions below the detectability threshold, while the outgoing proton can remain visible. This effect is significantly more pronounced in FASER$\nu$2 than in FLArE. A detailed treatment of this background will also depend on the position of the interaction in the tungsten layer, which we leave for future studies with more detailed detector simulations.

For completeness, we also present in \cref{tab:elasticbackgrounds} the number of DM signal events obtained for a benchmark scenario with $m_\chi = m_{A^\prime}/3 = 100~\mev$, $\varepsilon=6\times 10^{-4}$ ($y = 2.2 \times 10^{-9}$), 
and $\alpha_D=0.5$ for three sets of cuts and different detectors. Both in this table and in the subsequent analysis, the number of DM signal events has been additionally rescaled by a finite signal detection efficiency. This is due to the impact of FSI on the outgoing proton that can affect the DM-induced event reconstruction in the detector. We have estimated this efficiency as a function of the momentum of the final-state proton produced in the initial interaction inside the nucleus by studying elastic scatterings of neutrinos with \texttt{GENIE}. The value of the signal efficiency factor that we use in our analysis typically varies between $50\%$ and $70\%$, and it depends on the energy of the outgoing proton and the analysis type. As can be seen, for FLArE-10 and FLArE-100 with the lower limit $E_{k,p}>20~\mev$, the DM signal can yield a 30\% to 40\% excess over the neutrino background. In contrast, for FASER$\nu$2, even though the DM scattering rate is somewhat larger than in FLArE-10, the prospects for probing DM are limited by larger backgrounds.

\begin{figure}
    \centering
    \includegraphics[width=0.49\textwidth]{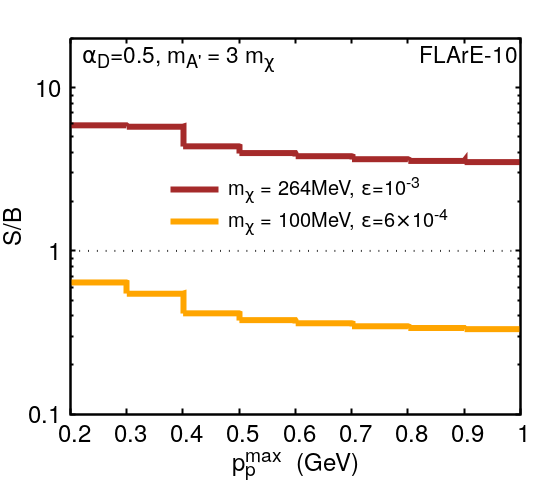}
\hfill
    \includegraphics[width=0.49\textwidth]{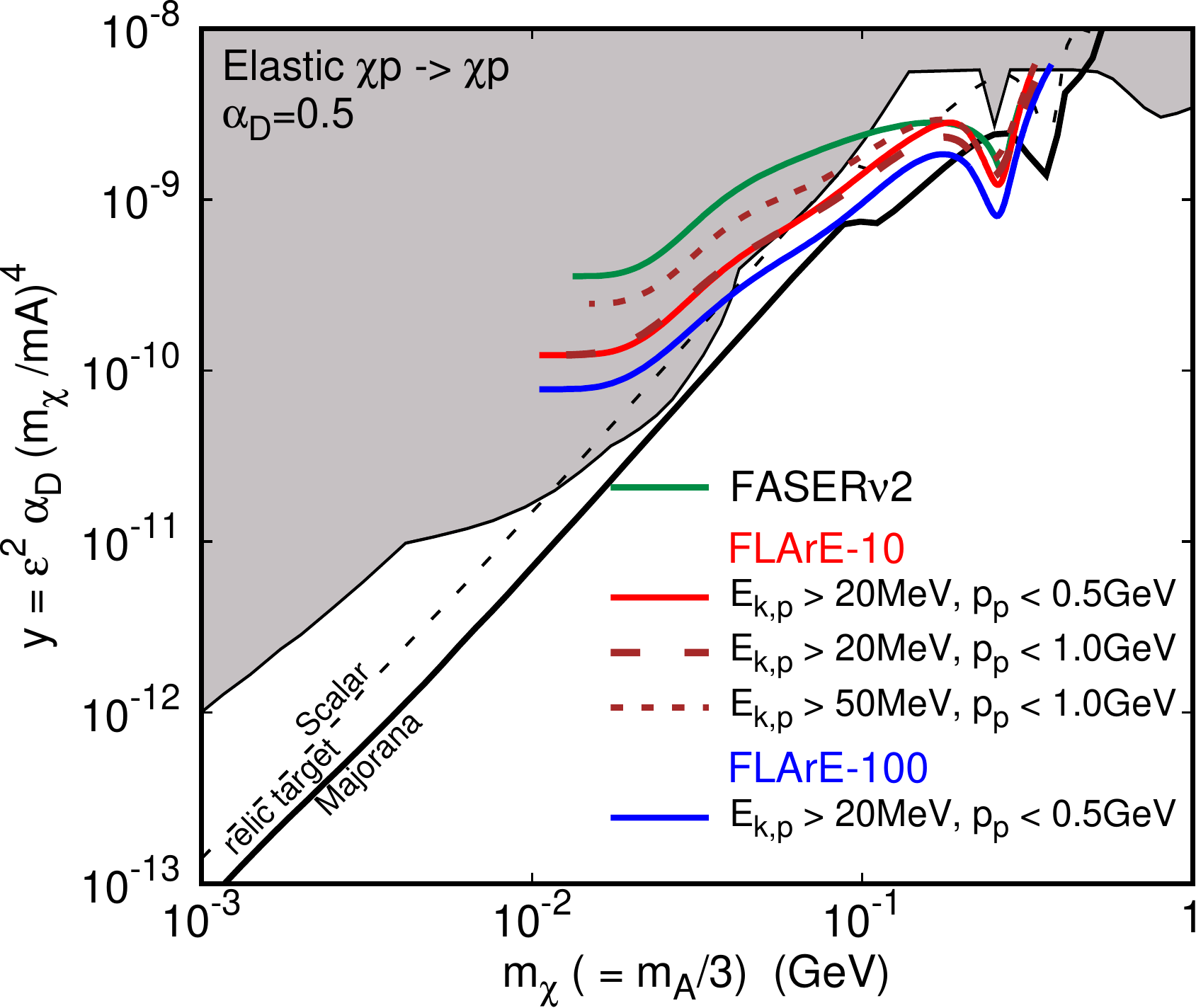}
\caption{\textsl{Left}: The signal-to-background ratio $S/B$ for the elastic scattering signature for FLArE-10 and the two DM benchmark scenarios indicated as a function of the maximum momentum of the outgoing proton $p_p^{\textrm{max}}$. The expected number of neutrino-induced background events for selected values of $p_p^{\textrm{max}}$ can be found in \cref{tab:elasticbackgrounds}, and we assume the detectability threshold of $E_{k,p}>20~\mev$ for the proton kinetic energy. \textsl{Right}: The projected $90\%$ CL exclusion bounds for the elastic scattering signature for FASER$\nu$2 with $300~\mev \alt p_p \alt 1~\gev$ (green), FLArE-10 (red), and FLArE-100 (blue) with the proton energy and momentum cuts indicated. Current bounds exclude the gray-shaded region; see \secref{combined} for details. The thermal relic targets for the Majorana fermion DM and complex scalar DM models are also shown.
}
\label{fig:elastic}
\end{figure}

In the left panel of \cref{fig:elastic}, we show the signal-to-background ratio $S/B$ as a function of $p_p^{\textrm{max}}$ for the FLArE-10 detector. We present results for the above-mentioned benchmark scenario and also for one with $(m_{\chi}, \varepsilon) = (264~\mev, 10^{-3})$ ($y = 6.2 \times 10^{-9}$). As evident from Fig.~\ref{fig:elastic}, the DM search favors lower values of $p_p^{\textrm{max}}$. This is expected for DM scatterings mediated by the dark photon $A^\prime$, which is much lighter than the $Z$ boson mediating neutrino NC scatterings. For a similar discussion for FLArE and DM-electron scattering, see Ref.~\cite{Batell:2021blf}. As is apparent from \eqref{dffxsecDMQE-Majorana}, the lower the $A'$ mass, the lower the typical momentum exchange in the $\chi p \rightarrow \chi  p $ reaction, which also leads to a lower characteristic momentum of the outgoing proton. In particular, for $m_{A^\prime}\lesssim 100~\mev$, it would become necessary to require $p_p^{\textrm{max}}\lesssim 300~\mev$ or even lower to obtain $S/B\sim 1$. This, however, goes beyond the FLArE and FASER$\nu$2 capabilities that we assume in our study. On the other hand, the DM scattering rate can become much higher for increasing mediator mass, in which case a larger momentum exchange is allowed. This can be seen for the case of $m_{A^\prime}= 3 m_\chi = 792~\mev$ also shown in the plot. The surprisingly large values of $S/B$ obtained for this benchmark scenario are related to the efficient $A^\prime$ production in the proton bremsstrahlung process for $m_{A^\prime}$ close to the $\rho$ and $\omega$ resonances. 

Last but not least, we note that, if systematic uncertainties are negligible relative to statistical uncertainties, the significance of the signal is more closely characterized by $S/\sqrt{B}$ than $S/B$.  As $p_p^\textrm{max}$ increases, the background rate increases, but this increase is milder for $\sqrt{B}$ than for $B$, and the dependence on the maximum momentum cut is milder for the ratio $S/\sqrt{B}$ than for $S/B$. For this reason, the projected exclusion bounds shown below are roughly independent of the precise value of $p_p^\textrm{max}$.

\subsection{Sensitivity Reach} 

In the right panel of \figref{elastic}, we present the expected projected $90\%$ CL exclusion bounds for the three detectors under study. 
We see that, with just the elastic scattering signature, FLArE-10 will probe most of the thermal relic target for the complex scalar DM model with $m_\chi \agt 100~\mev$. For the Majorana fermion DM case, FLArE-10 will only probe the small part of the thermal target region where DM production is enhanced by $\omega$ and $\rho$ resonances in the dark photon bremsstrahlung process. The detection prospects could be further improved in the larger FLArE-100 experiment. The expected exclusion bounds for FASER$\nu$2 are similar to FLArE-10.  We reiterate, however, that, as noted in \secref{detect}, this assumes that muon-induced backgrounds can be eliminated for FASER$\nu$2.

We also show the impact of different cuts on the proton kinetic energy, $E_{k, p} > 50~\mev$, and maximum outgoing proton momentum, $p_p^{\textrm{max}} < 1~\gev$. We see that the reach is better in the low-mass region for the lower proton kinetic energy threshold. However, the improved reach mainly corresponds to a region in the parameter space that is already excluded by existing bounds. On the other hand, the expected bound at higher masses is only slightly sensitive to changes of our lower kinetic energy and upper momentum cuts. As a result, the presented sensitivity reach for $m_\chi\gtrsim 100~\mev$ only mildly depends on the final FLArE capabilities in the considered range of $E_{k,p}$ and $p_p$. When we present combined results for different types of searches in \secref{combined}, we will therefore just present results for the cuts $E_{k, p} > 20~\mev, p_p < 0.5~\gev$ .

\section{Resonant Pion Production \label{sec:pion}}

\subsection{Signal} 

The next signal of interest is $\chi1\pi^0$ events, in which a single neutral pion is produced through DM-nucleus scattering with no other mesons or charged leptons emerging from the vertex. Such events are produced by DM-induced resonant pion production, $\chi N\to \chi N \pi^0$, which we model using the \texttt{BdNMC} DM simulation tool~\cite{deNiverville:2016rqh}. \texttt{BdNMC} accounts for incoherent pion production via excitation of the $\Delta$ resonance, which is expected to be the leading contributor to this process. In addition, $\chi1\pi^0$ events can also result from DM elastic scatterings off protons followed by FSI. We include this effect in our analysis, although it only mildly affects our final results.  When treating the elastic scattering contribution, we assume that the impact of FSI can be modeled using neutrino interactions, as was discussed in \secref{elastic}.

In our analysis, we do not differentiate events based on the number of final-state nucleons, including protons, that emerge from the nucleus. This is to mitigate the strong dependence of the number of expected signal events on the assumed FSI model. This inclusive approach is consistent with similar analyses performed by the K2K~\cite{Nakayama:2004dp}, MicroBooNE~\cite{MicroBooNE_osti_1573043} and MiniBooNE~\cite{AguilarArevalo:2009ww} Collaborations. 

\begin{figure}
    \centering
    \includegraphics[width=0.48\textwidth]{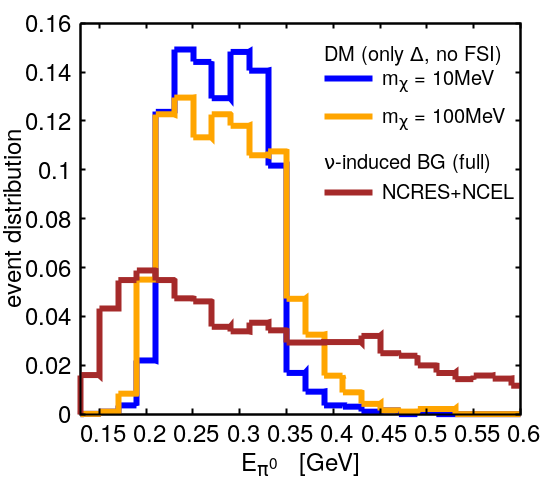}
\hfill
    \includegraphics[width=0.49\textwidth]{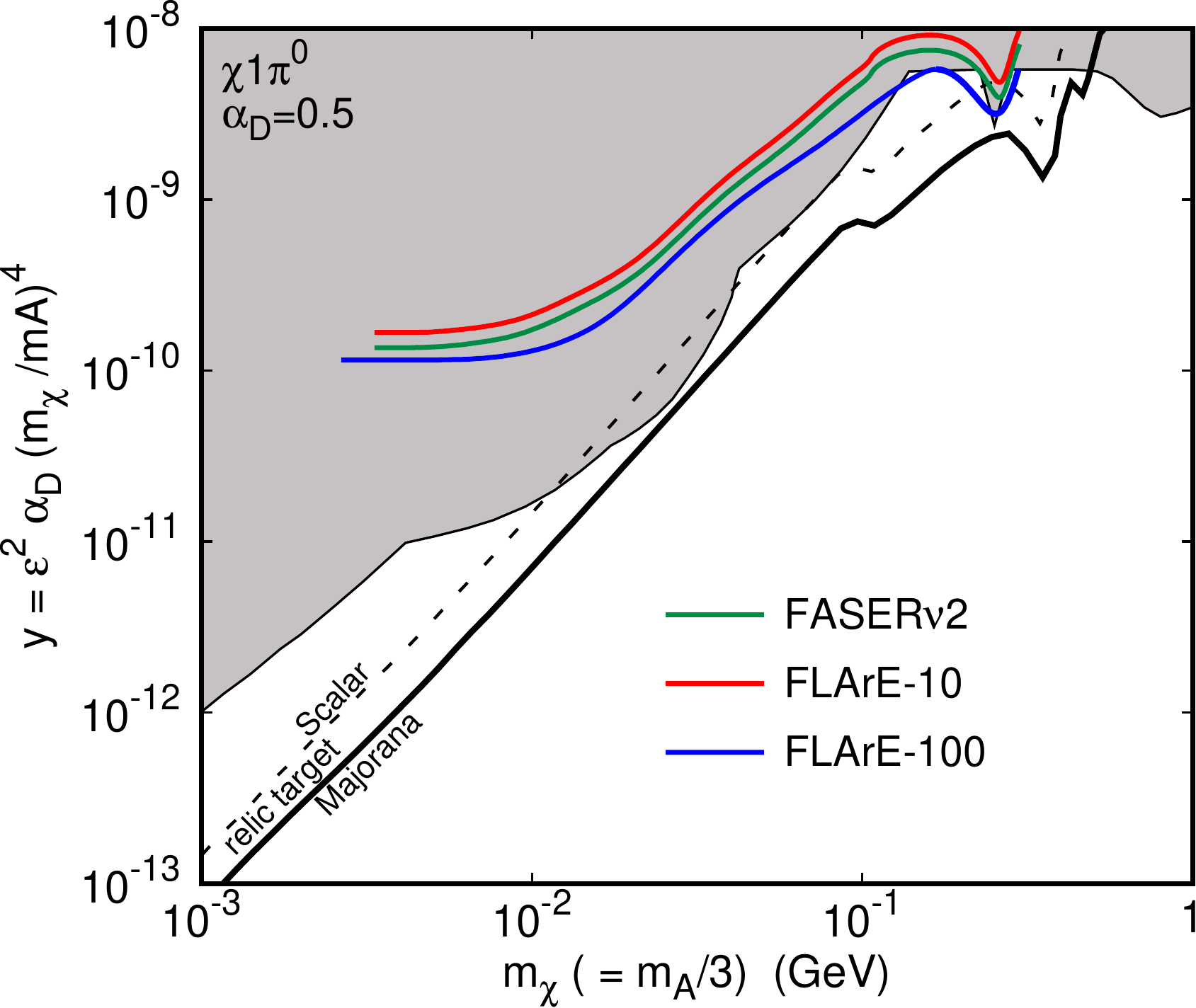}
\caption{\textsl{Left}: The event distribution as a function of the pion energy for $\chi1\pi^0$ signal events and neutrino-induced backgrounds in the liquid argon detectors. The DM results are shown for two benchmark masses $m_\chi = m_{A^\prime}/3 = 10~\mev$ (blue) and $100~\mev$ (yellow) for the complex scalar DM model. They have been obtained with the \texttt{BdNMC} code~\cite{deNiverville:2016rqh} that takes into account the dominant pion production via production of the $\Delta$ resonance. We also show the relevant results for neutrino-induced backgrounds from NCRES and NCEL events (brown histogram). This was obtained using the far-forward LHC neutrino energy spectrum and full \texttt{GENIE}~\cite{Andreopoulos:2009rq,Andreopoulos:2015wxa} simulations with further resonances and final-state interactions of hadrons taken into account. \textsl{Right}: The colorful solid lines correspond to the projected $90\%$ CL exclusion bounds in the DM-nuclei scattering $\chi1\pi^0$ signature for FASER$\nu$2 (green), FLArE-10 (red), and FLArE-100 (blue). Current bounds and thermal relic targets are as in \figref{elastic}.   
\label{fig:resonant}}
\end{figure}

The neutral pion in the final state will immediately decay into two photons with momenta typically above the visibility threshold of $30~\mev$ characteristic for the liquid-argon detectors. In contrast, for FASER$\nu$2, the reach will partially be limited by the requirement that photons have an energy of at least $300~\mev$ to be visible. 
As discussed above, in the resonant pion production events, we typically observe $E_\pi\sim 300~\mev$ from the $\Delta$ resonance, which would only be moderately 
altered by the presence of heavier resonances and FSI. We illustrate this in the left panel of \figref{resonant}, in which we show the resonant event distribution as a function of the energy of the final-state neutral pion $E_{\pi^0}$ for two benchmark DM models with $m_\chi = m_{A^\prime}/3 = 10$ and $100~\mev$, and for neutrino-induced NCRES background events. The plot has been obtained for the liquid argon detector. As can be seen, in the case of neutrinos, in which the aforementioned effects going beyond the simple $\Delta$ resonance and parton level interactions are taken into account, the resulting distribution is more smeared than for DM. 
Notably, in both cases, the photons produced in $\pi^0$ decays will typically be above 
$30~\mev$.

The characteristic energy of the pions produced through resonant scatterings translates into a relatively weak dependence of the sensitivity reach on the upper energy threshold, which is similar to the elastic DM-nucleon scattering search discussed in \secref{elastic}. 
As a result, we will employ a single cut on the maximum pion energy given by $E_{\pi^0} < 1~\gev$. Increasing this limit has a minimal impact on the number of DM-induced resonant pion production events, while it could adversely affect the sensitivity by increasing the number of neutrino-induced backgrounds from DIS events. 

Similar to the discussion in \secref{elastic}, here we also do not discuss the possible impact of the angular cuts on the derived exclusion bounds. We note, however, that the pion angular distribution, as well as the invariant mass reconstruction of the photon pair, could play an important role in further distinguishing such events from neutrino-induced backgrounds producing single electrons in the final state due to the scatterings off electrons or nuclei; see Ref.~\cite{Aguilar-Arevalo:2018wea} for a similar discussion for MiniBooNE. Below, for simplicity, we assume that such backgrounds can be rejected in the analysis.

\subsection{Neutrino-Induced Backgrounds}

The dominant neutrino-induced backgrounds for the $\chi1\pi^0$ events are due to NCRES scatterings. We also study subdominant contributions associated with the coherent pion production processes (COHERENT), in which the neutrino scatters off the entire nucleus, and the elastic scatterings NCEL followed by the FSI that generate the outgoing neutral pion. We model all these backgrounds using \texttt{GENIE}. We provide the total expected number of background events for the three detectors in \tableref{resonantbackgrounds} for four choices of the $\pi^0$ upper energy threshold: $E_{\pi^0}< 170~\mev$, $300~\mev$, $1~\gev$, and $2~\gev$. As can be seen, increasing the energy threshold above $1~\gev$ has a very mild impact on the number of background events. We require that the events do not contain any charged pions or other mesons above the visibility thresholds discussed in \cref{sec:detectors}.

\begin{table}
\begin{center}
\begin{tabular}{ c||c|c|c|c } 
\hline
\hline
$\mathbf{\chi1\pi^0}$ & \multicolumn{4}{c}{$\nu$-induced backgrounds}\\
\cline{1-1}
Detector & $E_{\pi^0}<170~\mev$ & $300~\mev$ & $1~\gev$ & $2~\gev$\\
 \hline

 FASER$\nu$2 & -- & -- & $150$ & $170$\\
 FLArE-10 & $9$ & $90$ & $220$ & $230$\\ 
 FLArE-100 & $70$ & $740$ & $1750$ & $1850$\\ 
 \hline
 \hline
\end{tabular}
\end{center}
\caption{Neutrino-induced background events in the search for $\chi1\pi^0$-type events (see the text for details) as a function of the maximum threshold for the outgoing pion energy. The minimum threshold energy for the outgoing photon is set to $300~\mev$ and $30~\mev$ for the emulsion and liquid argon detectors, respectively.
\label{table:resonantbackgrounds}}
\end{table}

Focusing now on the pion energy cut of $E_{\pi^0}\lesssim 1~\gev$, we see that we expect roughly 200 background events in both FASER$\nu$2 and FLArE-10, and roughly 2000 such events in FLArE-100. Interestingly, the number of background events is now smaller in FASER$\nu$2 than for FLArE-10. This is the opposite effect to the one discussed in \secref{elastic}, in which increasing the lower energy threshold resulted in a larger number of NCRES events mimicking NCEL ones in the detector. For this reason, we now observe a relatively lower number of NCRES events that will be reconstructed in the emulsion detector as $\chi1\pi^0$-like events. As far as liquid argon detectors are concerned, the number of background events in this search is larger, although of a similar order, than for the previously discussed search based on elastic scattering events. 

\subsection{Sensitivity Reach}

In the right panel of \figref{resonant}, we present the expected projected $90\%$ CL exclusion bounds based on the $\chi1\pi^0$ search. As can be seen, the expected bounds are weaker than the ones based on DM elastic scattering shown in \figref{elastic}. This is primarily due to the smaller scattering cross section. Once we limit the DM signal rate to only NC ($A^\prime$ exchange) scatterings off protons with single $\pi^0$ production and no charged pions in the final state, the relevant cross section is suppressed relative to elastic scattering by more than an order of magnitude for small mediator masses, $m_{A^\prime}\lesssim 100~\mev$~\cite{deNiverville:2016rqh}. The suppression factor becomes smaller, of order a factor of a few, for heavier dark photons. The signal rate is further suppressed by signal efficiencies resulting from FSI and event reconstruction. We estimate them to be of the order of $25\%$ for FLArE and between $10\%$ and $15\%$ for FASER$\nu$2. In the latter case, this efficiency also takes into account the aforementioned energy cut of $E_{\gamma}\gtrsim 300~\mev$, which is larger than in LArTPC detectors. In the end, we find that the resonant pion signature is less promising than both the electron and single proton signatures.

\section{Deep Inelastic Scattering\label{sec:dis}}

\subsection{Signal}

The last signature that we consider is DM-nuclear scattering at high momentum transfer. Because light DM will be produced with TeV-scale energies in the direction of the FPF, the maximum accessible momentum transfer in nuclear scattering is tens of GeV. Above the QCD scale, deep inelastic scattering leads to a relatively high-energy nuclear recoil, which can subsequently produce multiple charged tracks. In this regime, a partonic treatment is appropriate, and the outgoing hadrons are easily above detector thresholds.

We consider the DIS process $\chi N \to \chi X$ in the models of \cref{sec:modelswithAprime}. The double differential cross section is given by 
\begin{equation}
\frac{d\sigma(\chi N \!\to\! \chi X)}{dx \ dy} 
= 2 \pi \varepsilon^2 \alpha \alpha_D  \frac{2 m_p E_\chi}{(Q^2+m_{A'}^2)^2}  \!\!\!\! \sum_{q=u,d,s,c}\!\!  Q^2_q \ B(y) \big[ x f_{q}(x,Q^2) + x f_{\bar{q}}(x,Q^2) \big ] \, ,
\end{equation}
where $Q^2 = 2 m_p E_\chi x y$, $x$ is the parton momentum fraction, $y = 1 - E'_{\chi} / E_{\chi}$ is the fraction of the incoming DM energy transferred to the nucleon in the lab frame, $f_q$ is the quark parton distribution function, $Q_q$ is the quark electric charge, and 
\begin{equation}
   B(y) = 
   \begin{cases}
 \displaystyle{ 1 + (1 - y)^2}  \quad \text{(Majorana fermion DM)}
   \\
\displaystyle{ 2(1 - y) } \quad \text{(complex scalar DM)} \ .
   \end{cases} 
\label{eq:Bfactor}
\end{equation}
As the scattering takes place through a light mediator, it is not surprising that low momentum transfer is favored regardless of the $\chi$ spin. Furthermore, the functions $B(y)$ for Majorana fermion and complex scalar DM in \eqref{Bfactor} are identical up to $\mathcal{O}(y^2)$. Because the cross section is dominated by the small $y$ region, then, the DIS signal strength is approximately the same for these two models. This motivates the choice previously mentioned in Sec.~\ref{sec:modelswithAprime} to only show results for the Majorana fermion DM scenario.

To estimate the scattering signal, we convolute these cross sections with the nCTEQ15 parton distribution functions~\cite{Kovarik:2015cma} for tungsten and argon nuclei, imposing a minimum cut of $Q^2 > 1~\mathrm{GeV}^2$. When the parton hadronizes, of course, multiple charged tracks and photons, which yield electromagnetic showers, are produced. We do not simulate this hadronization nor the reconstruction of the hadronic energy and transverse momentum from these objects, though other works have demonstrated the use of track-level information to search for similar signals~\cite{Machado:2020yxl, Ismail:2020yqc}. Instead, we simply take the outgoing parton energy and transverse momentum as proxies for the energy and transverse momentum of the recoiling hadronic system, 
\begin{eqnarray}
E_\text{had} = y E_\chi 
\quad \text{and} \quad
p_{T,\text{had}}^2 = Q^2 (1 - y) = 2 m_p E_\chi x y (1 - y) \ .
\label{eq:disepthad}
\end{eqnarray}
We expect both $E_\text{had}$ and $p_{T,\text{had}}$ to grow with increasing $Q^2$. In principle, there are more detailed kinematic variables involving the visible tracks from the scattered nucleon that could be accessed by doing a full simulation. However, since the hadronic part of each event depends only on the outgoing parton momentum and hadron interaction modeling, we do not anticipate that further kinematic considerations would provide significant additional discriminating power between the signal and neutrino background.

\begin{figure}
    \centering
    \includegraphics[width=0.99\textwidth]{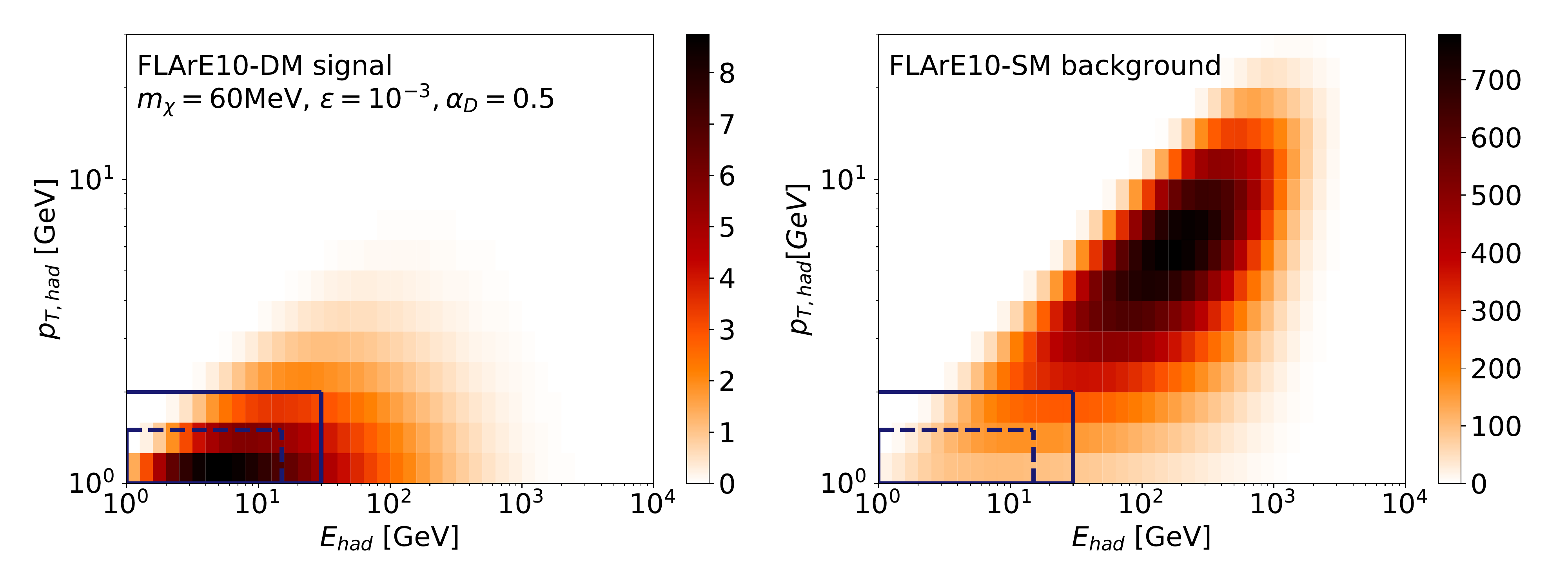}
    \caption{
    Expected number of DIS events in the $(E_{\text{had}}, p_{T,\text{had}})$ plane for one benchmark Majorana DM scenario (left) and SM NC neutrino background (right) at FLArE-10. Most of the signal events are at low $E_{\text{had}}$ and low $p_{T,\text{had}}$, motivating our choice of cuts. The dashed (solid) box shows the strong (loose) cuts of 1 GeV $<E_{\text{had}}<$ 15 (30) GeV and 1 GeV $<p_{T,\text{had}}<$ 1.5 (2.0) GeV used in our analysis. 
    }
    \label{fig:DIS2D_majorana_DM}
\end{figure}

The left panel of \figref{DIS2D_majorana_DM} shows the two-dimensional distribution of the quantities in \eqref{disepthad} for DIS in one of our benchmark DM scenarios at FLArE-10. The distribution is qualitatively similar at FASER$\nu$2. The signal events are clustered at lower energies and transverse momenta than the background, consistent with the preference for low momentum transfer in light DM scattering. Despite the preference for low momentum transfer, there is still a significant number of events with energetic nuclear recoils. We see the most events at $E_\text{had}$ of several GeV and low $p_{T,\text{had}}$, and expect that such events would have multiple tracks emerging from a vertex with no incoming track. A more detailed study of the detection efficiency, including the effects of hadronization and FSI, would be interesting. For instance, the efficiency would depend on the number of tracks and hence the hadron multiplicity, which tends to grow with the center-of-mass energy $W$ of the recoiling hadronic system. $W$ is related to the momentum transfer and partonic momentum fraction through $W^2 = m_p^2 + Q^2 (1 - x) / x$. The EMC experiment measured the charged hadron multiplicity in muon DIS, finding that several charged tracks were typical for $W > 4~\gev$~\cite{Arneodo:1985xr}. We have checked that a cut of $W > 2~\gev$, which would avoid the resonant scattering region with fewer tracks, does not change our results significantly. In addition, as our signal is clustered at values of $p_{T,\text{had}} / E_\text{had}$ corresponding to angles of several degrees, it would be useful to examine technologies for measuring multiple hadronic tracks in the forward direction in liquid argon for FLArE. While there can be difficulties measuring such tracks using wire planes if the planes are oriented parallel to the track direction, the patterns of charge deposition can be used to obtain three-dimensional information~\cite{Qian:2018qbv}, as has been demonstrated by MicroBooNE for neutrino event identification~\cite{Abratenko:2020hpp} and cosmic ray rejection~\cite{Abratenko:2021bzb}.

\subsection{Neutrino-Induced Backgrounds}

The main background to DM DIS is neutrino scattering. NC neutrino scattering would produce a nuclear recoil with significant energy carried away by the outgoing neutrino, just as in our signal. CC neutrino scattering, by contrast, would result in a high-energy outgoing lepton. We assume that the detector would have sufficient efficiency that the neutrino CC backgrounds could be rendered very small.

There are also backgrounds from muon interactions, which can be eliminated by requiring that there is no charged track leading into the vertex~\cite{Abreu:2019yak}. Muon interactions can also produce neutral hadrons, which travel for distances on the order of 10 cm before scattering. These neutral hadron events can mimic the signal. Although neutral hadron backgrounds are problematic in a pure emulsion detector~\cite{Abreu:2019yak,Ismail:2020yqc}, as mentioned in \secref{detect}, we assume that an active muon veto will remove these events at FASER$\nu$2 or FLArE~\cite{Batell:2021blf}. By using timing to remove a small area around each muon interaction, we expect that neutral hadron scattering could be reduced to negligible levels without significant impact on the signal.

The differential NC neutrino scattering cross section at high energy is~\cite{McFarland:2008xd}
\be
    \frac{d\sigma(\nu N \to \nu X)}{dx \ dy} = \frac{ 2G_F^2 m_p E_\nu}{ \pi} \frac{m_{Z}^4}{(Q^2+m_{Z}^2)^2} 
    \times  \!\!\!\! \sum_{q=u,d,s,c} \!\!&\big[ g_{q,L}^2 [x f_{q}(x,Q^2) + x f_{\bar{q}}(x,Q^2)(1-y)^2 ] \\
    & + g_{q,R}^2 [x f_{q}(x,Q^2) (1-y)^2 + x f_{\bar{q}}(x,Q^2) ] \big ]
\ee
in terms of the partonic momentum fraction $x$ and the fractional neutrino energy loss $y = 1 - E'_{\nu} / E_{\nu} = E_\text{had}/E_\nu$. The momentum transfer is $Q^2 = 2 m_p E_\nu x y$. Here, $g_{q,L}, g_{q,R} = T^3 - Q \sin^2 \theta_W$ are the NC couplings of the quarks. For anti-neutrinos, the cross section is 
\be
    \frac{d\sigma(\bar{\nu} N \to \bar{\nu} X)}{dx \ dy} = \frac{ 2G_F^2 m_p E_\nu}{ \pi} \frac{m_{Z}^4}{(Q^2+m_{Z}^2)^2} 
    \times  \!\!\!\! \sum_{q=u,d,s,c} \!\!&\big[ g_{q,L}^2 [x f_{q}(x,Q^2)(1-y)^2 + x f_{\bar{q}}(x,Q^2)] \\
    & + g_{q,R}^2 [x f_{q}(x,Q^2) + x f_{\bar{q}}(x,Q^2) (1-y)^2] \big ] .
\ee
As the momentum transfer $Q^2$ is generally small compared to $m_Z^2$, the neutrino scattering cross sections are proportional to the CM energy or, equivalently, the energy of the incoming neutrino.

The typical $Q^2$ is perhaps the most striking difference between light DM DIS and neutrino NC scattering. In principle, the momentum transfer $2 E_\chi m_N$ in DM scattering can be as high as tens of GeV. However, for scattering through a light mediator, smaller momentum transfers are typically preferred, as the scattering cross section goes as $1/Q^4$ in the limit of vanishing mediator mass. On the other hand, neutrino scattering proceeds through the $Z$, which is heavy compared to the typical momentum transfer. Consequently, the neutrino NC scattering cross section grows linearly with the partonic CM energy $\sqrt{\hat{s}}$.

We proceed to investigate the kinematics further to discriminate between signal and background, showing the hadronic energy and transverse momentum for the neutrino background in the right panel of Fig.~\ref{fig:DIS2D_majorana_DM}. Motivated by these kinematic distributions, we consider two sets of cuts on $E_\text{had}$ and $p_{T,\text{had}}$:
\be
    \text{Strong cuts:} && 1~\gev<E_\text{had}<15~\gev \, ,&& \ \ 1~\gev<p_\text{T,had}<1.5~\gev && \\
    \text{Loose cuts:} && 1~\gev<E_\text{had}<30~\gev \, ,&& \ \ 1~\gev<p_\text{T,had}<2.0~\gev && .
    \label{eq:discuts}
\ee
The effects of these cuts on the background and signal are shown in \cref{tab:dis}. We see that the background can be reduced by over an order of magnitude while keeping 1/4 to 1/2 of the DM DIS signal.

\begin{table}
\begin{center}
\begin{tabular}{ c||c|c|c||c|c|c||c|c|c } 
\hline
\hline
$\mathbf{DIS}$ & \multicolumn{3}{c||}{$\nu$-induced backgrounds} &
\multicolumn{3}{c||}{DM: $m_\chi=60~\mev, \varepsilon=10^{-3}$} &
\multicolumn{3}{c}{DM: $m_\chi=188~\mev, \varepsilon=10^{-3}$}\\
\cline{1-1}
Detector 
& no cuts & loose cuts & strong cuts 
& no cuts & loose cuts & strong cuts 
& no cuts & loose cuts & strong cuts \\
 \hline
 FASER$\nu$2 & 154k & 7.4k  & 2.9k &
               700 & 335 & 210 & 
               440 & 170 & 100  
 \\ 
 FLArE-10    & 82k & 5k & 2k &
               380 & 185 & 116 &
               250 & 95 & 55  
 \\ 
 FLArE-100   & 528k & 38k & 15k &
               2.3k   & 1.1k & 748 &
               1.5k & 615 & 361  
 \\ 
 \hline
 \hline
\end{tabular}
\end{center}
\caption{The effects of the energy and momentum cuts in \eqref{discuts} on the numbers of SM neutrino NC background and DM DIS signal events. Two different benchmark DM scenarios are shown. The ``no cuts'' columns include only a $Q^2$ requirement and no cuts on the hadronic transverse momentum or energy.}
\label{tab:dis}
\end{table}

\subsection{Sensitivity Reach}

\begin{figure}
    \centering
    \includegraphics[width=0.99\textwidth]{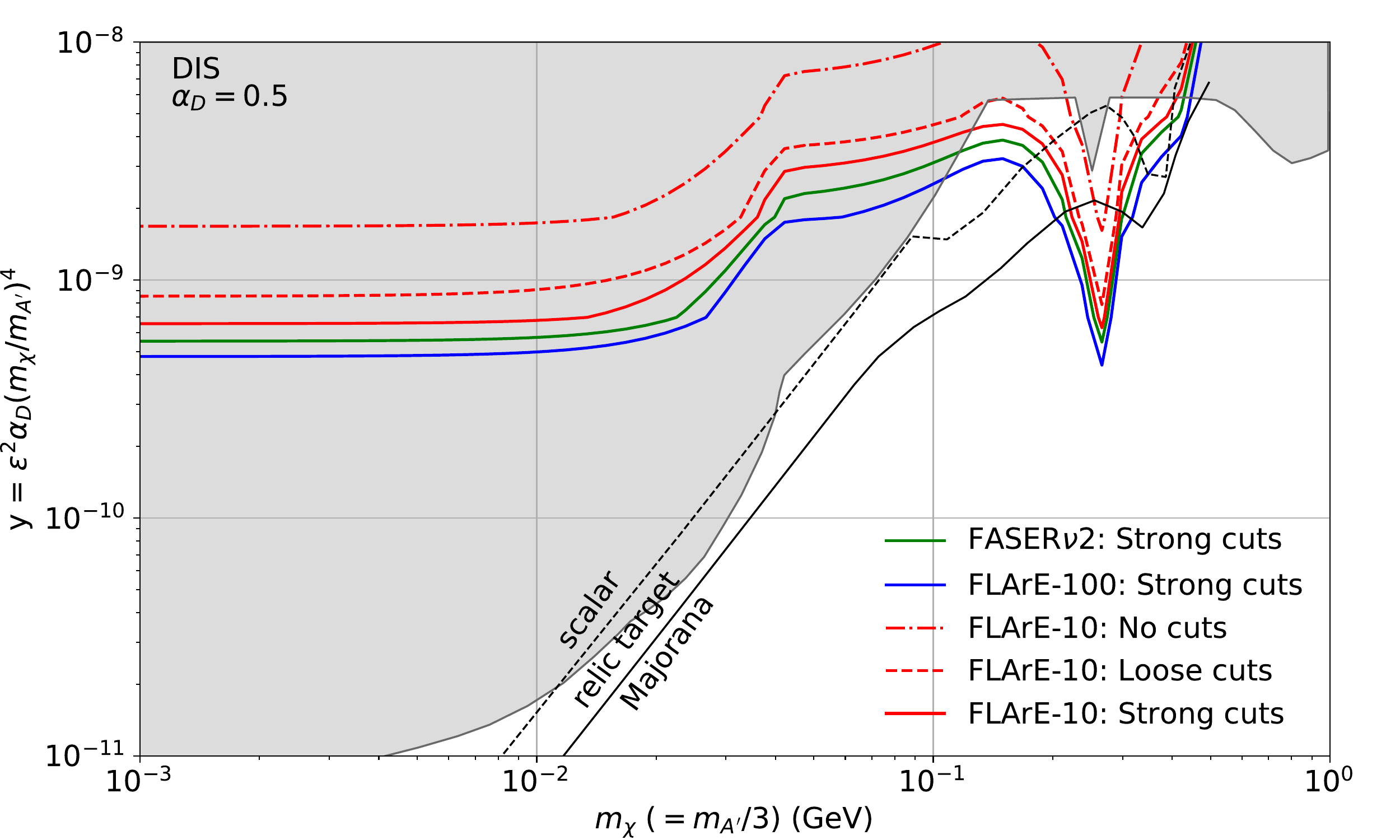}
    \caption{The projected 90\% CL exclusion bounds for the DIS signature in the Majorana fermion DM model at various detectors. For FLArE-10 we show the limits with and without the kinematic cuts, whereas for FASER$\nu$2 and FLArE-100 we show only the best limits corresponding to the strong cuts. The thermal relic targets for Majorana fermion DM (black solid) and complex scalar DM (black dashed), and current bounds (gray shaded region) are also shown. 
    }
    \label{fig:DISexclusionplot}
\end{figure}

Having examined the kinematics of the signal and background events, we present the expected projected $90\%$ CL exclusion bounds for DM DIS searches at FASER$\nu$2 and FLArE in \figref{DISexclusionplot}. Considering statistical uncertainties only, the former set of cuts in \eqref{discuts} yields the strongest projected exclusions. The figure shows the reach of the different detectors, as well as the effects of the hadronic energy and transverse momentum cuts in the case of FLArE-10. In contrast to the lower energy signatures in Secs.~\ref{sec:elastic} and \ref{sec:pion}, the typical deposited energy is well above the thresholds for both emulsion and liquid argon detectors. The relative performances of FASER$\nu$2 and FLArE thus depend mostly on the detector mass and geometry, as well as on their background rejection and event identification capabilities. Here, we focus on the former, while assuming $100\%$ signal detection efficiency for both types of experiments. Of the two 10-tonne detectors, the more compact FASER$\nu$2 provides better sensitivity to light DM scattering because it has more mass at large rapidity where the DM flux is higher.
In addition, the numbers of events for FLArE-100 in Table~\ref{tab:dis} do not scale fully with the detector mass, when compared to its 10-tonne analog. Similar effects were observed for DM-electron scattering~\cite{Batell:2021blf}.

As discussed above, the DIS limits are very similar for the Majorana fermion and complex scalar DM models, and we have used the former to draw the projected exclusion lines. To guide the interpretation of the limits, we also show the thermal relic targets in each of these scenarios, assuming standard thermal cosmology. We see that DIS searches can probe dark photon scenarios yielding the correct thermal relic density for DM masses above approximately 200 MeV. The expected sensitivity reach can then also partially cover the resonance region, in which the intermediate dark gauge boson in DM annihilations mixes with the SM vector mesons $\rho$ and $\omega$, i.e., $2 m_\chi \approx m_{\rho,\omega}$, especially for complex scalar DM. By contrast, the reach of DM DIS is relatively limited at low masses. This is because the growth of the DIS cross section at small mediator masses is limited by our minimum momentum transfer cut of 1 GeV. Nevertheless, DM DIS searches at FPF detectors offer the potential to probe dark photon scenarios that are viable from the standpoint of thermal cosmology and that are otherwise unconstrained.

Finally, we note from Table~\ref{tab:dis} that with the full HL-LHC dataset, there will be thousands of background events even with kinematic cuts. It will thus be important to reduce uncertainties from systematics such as the neutrino flux and signal/background modeling, which we have not considered here, in a full experimental analysis. We assume that they can be suppressed so that the analysis will be dominated by statistical uncertainties. For instance, as has been suggested previously~\cite{Batell:2021blf}, measuring the neutrino flux at other detectors or in other kinematic regions could help constrain the background normalization. If statistical uncertainties dominate, then since the number of signal events scales with $y^2$, the limit on $y$ associated with a fixed significance $S/\sqrt{B}$ improves as $\mathcal{L}^{-1/4}$. The impact of this mild dependence is that new parameter space can be probed with a relatively small amount of data. We will consider the effect of luminosity on the reach more completely in the next section, where we combine the results of this section and the previous two to obtain the overall FPF reach in searches for light DM-nuclear scattering.

\section{Combined Sensitivity Reach\label{sec:combined}}

In this section, we combine all of our previous results on DM-nucleus scattering processes, including elastic scattering (\cref{sec:elastic}), resonant pion production (\cref{sec:pion}), and DIS (\cref{sec:dis}), as well as the results previously obtained~\cite{Batell:2021blf} for the DM search based on scatterings off the electrons.

\begin{figure}
    \centering
\includegraphics[width=0.99\textwidth]{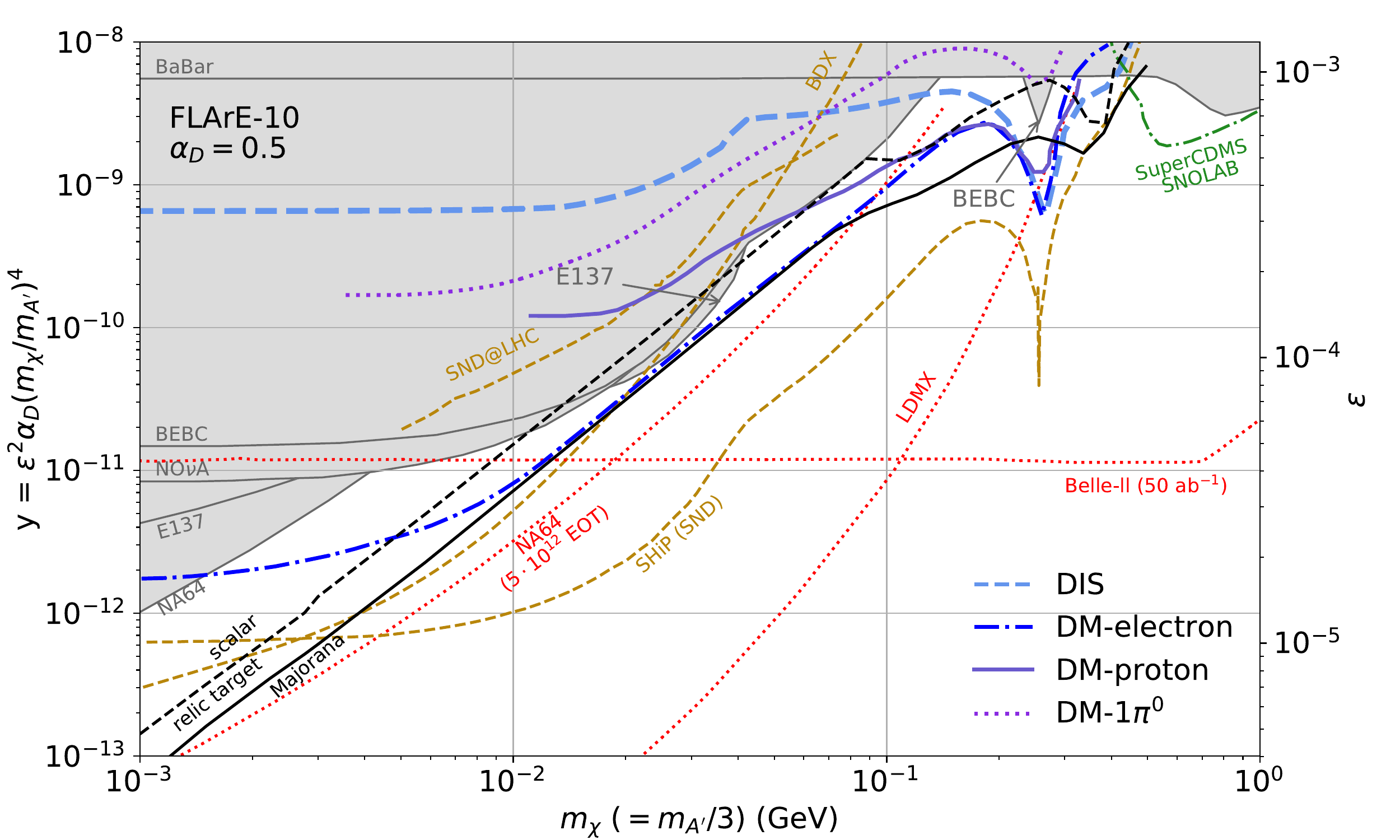}
    \caption{The projected 90\% CL exclusion bounds for Majorana fermion DM from DM-nucleus elastic scattering, resonant pion production, and DIS (this work), along with DM-electron scattering from Ref.~\cite{Batell:2021blf} at  FLArE-10. In the gray shaded region, we also show the strongest existing constraints from BaBar, NA64, NO$\nu$A, E137, and BEBC, as implemented in Refs.~\cite{Buonocore:2019esg, Aguilar-Arevalo:2018wea}. Projected reaches from other experiments are shown in brown for beam dump/collider experiments and in red for missing momentum-type searches. The green contour shows the projected bound on Majorana fermion DM from SuperCDMS; see text for more details.}
    \label{fig:FL10combined1}
\end{figure}

\begin{figure}
    \centering
\includegraphics[width=0.99\textwidth]{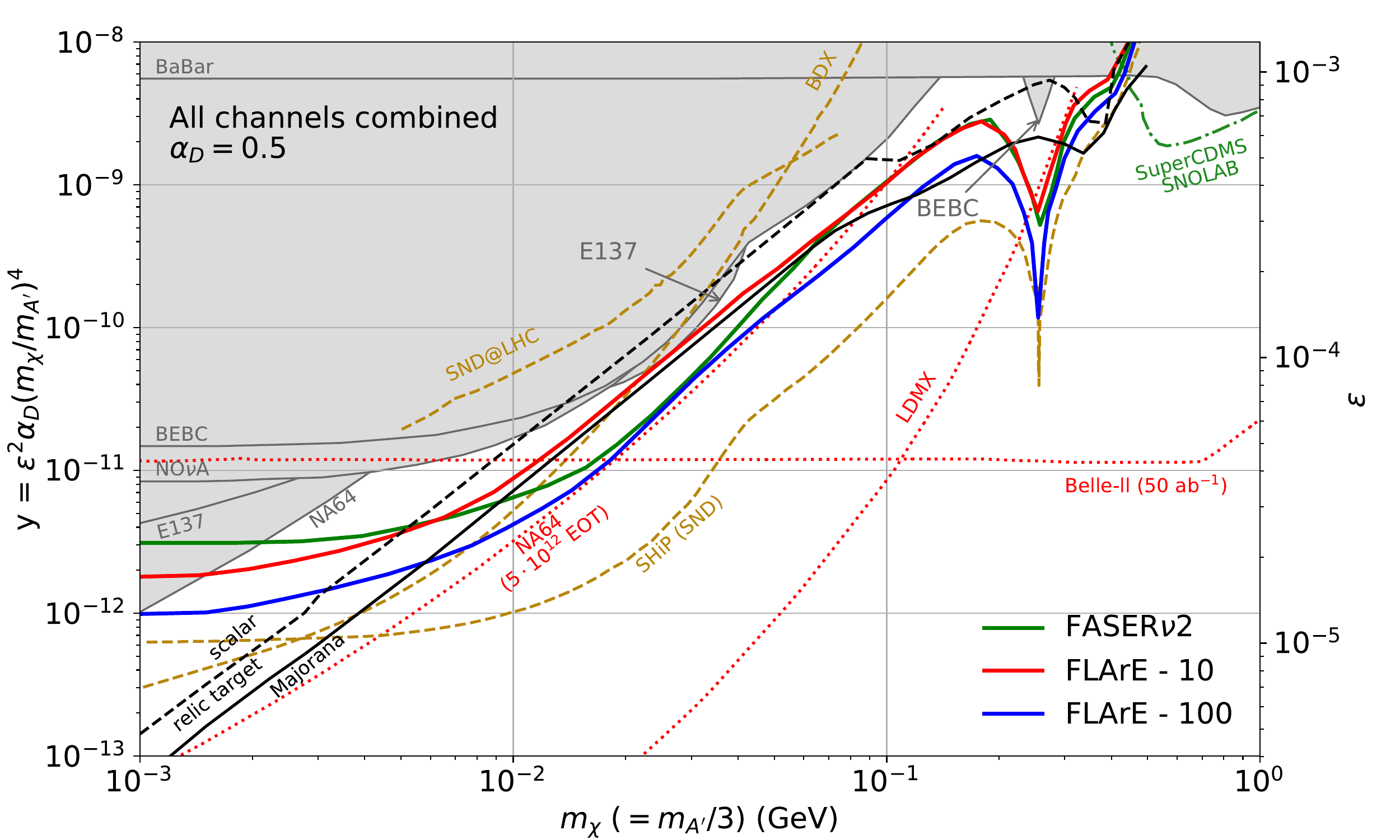}
 \caption{The projected 90\% CL exclusion bounds combining all channels for the FASER$\nu$2, FLArE-10, and FLArE-100 detectors at the HL-LHC with 3 ab$^{-1}$ of integrated luminosity. At lower DM mass the DM-electron signature is the best, whereas, at higher masses, DIS provides the most stringent limits. Existing constraints and projected reaches from other experiments are as in \figref{FL10combined1}.}
 \label{fig:FL10combined2}

\end{figure}

These are shown for FLArE-10 in \figref{FL10combined1}.  In general, since the scattering cross sections grow for small mediator mass and we have taken a fixed mass ratio $m_{A'} / m_\chi$, the limits are strongest at low $m_\chi$. The flattened sensitivity at the left of the plot arises from the minimum momentum transfer for each process considered. For elastic scattering and resonant production, these come from experimental considerations on the visibility of the outgoing proton or pion. We see that the low thresholds of liquid argon detectors allow for the ability to probe new parameter space at $m_\chi \lesssim 200~\mathrm{MeV}$. For DIS, the $Q^2$ cutoff to ensure the validity of our partonic treatment limits the sensitivity at small masses, but the inherently harder nature of DIS can lead to stronger bounds at higher DM mass.

\Figref{FL10combined1} also shows that elastic scattering and DIS are the most sensitive nuclear scattering probes at low and high masses, respectively. Resonant pion production is never the strongest signature in this model. The sensitivity reach from DM-electron scattering, derived previously in Ref.~\cite{Batell:2021blf}, is also shown, and can be seen to be competitive with the best nuclear signatures at moderate and high masses, and even stronger at low masses.

In \figref{FL10combined1}, we also show the thermal relic targets for Majorana fermion and complex scalar DM, as well as current and projected results from other experiments.  Existing bounds from null results are shown as the gray shaded region. These include results from BaBar~\cite{Lees:2017lec}, MiniBooNE~\cite{Aguilar-Arevalo:2018wea}, and NA64~\cite{NA64:2019imj}, as well as recasts of searches at BEBC~\cite{Grassler:1986vr}, CHARM-II~\cite{DeWinter:1989zg}, E137~\cite{Bjorken:1988as,Batell:2014mga}, LSND~\cite{deNiverville:2011it}, and NO$\nu$A~\cite{Wang:2017tmk}, as derived by the authors of Refs.~\cite{Marsicano:2018glj,Buonocore:2019esg}.  Projected sensitivities of future experiments are shown in the dashed and dotted colored contours.  We also note that future short baseline neutrino experiments such as ICARUS could also be sensitive to DM scattering~\cite{Buonocore:2019esg}. The brown contours are projected sensitivities from searches for DM that is produced at a collider or beam dump and then subsequently scatters in a downstream detector, a signature similar to what we have considered in this work. These include BDX~\cite{Battaglieri:2016ggd}, SND@LHC~\cite{Ahdida:2020evc}, and SND@SHiP~\cite{SHiP:2020noy}. The red contours are projected sensitivities of future missing momentum-type searches, including NA64~\cite{Gninenko:2019qiv}, LDMX~\cite{Berlin:2018bsc,Akesson:2018vlm}, and Belle-II~\cite{Kou:2018nap}. Last, the green contour shows the projected sensitivity of SuperCDMS to the Majorana fermion DM model~\cite{Battaglieri:2017aum,Akesson:2018vlm,Berlin:2018bsc}. The region probed by SuperCDMS is at higher masses than are probed by FLArE-10.  For the complex scalar DM model, direct detection limits can be more constraining, but they can also be evaded by the introduction of a small mass splitting between the DM states so that the scattering is inelastic. 

\Figref{FL10combined2} then shows the best limits from each of the detectors in \secref{detectors}. As for FLArE-10 in \figref{FL10combined1}, the best limits arise from electron scattering and nuclear DIS in the low and high mass ranges, respectively. Both FASER$\nu$2 and FLArE-10 will probe the relic target for the complex scalar DM model for DM masses between several MeV and a few hundred MeV. FLArE-100 could provide a similar reach for the Majorana fermion DM model. Altogether, the detectors we have studied are able to probe a wide swath of the cosmologically-favored parameter space for both the Majorana fermion and complex scalar DM models.

Finally, to estimate the time scales on which forward LHC searches could start to achieve new sensitivity to light DM, we show the 90\% projected exclusion bounds at FLArE-10 for a selection of integrated luminosities in \figref{combinedFLArE10lumi}. Again, the best limits from all processes (elastic proton scattering, resonant pion production, DIS, and electron scattering) have been used. With around $30~\ifb$ of data, these searches can begin to test thermal DM scenarios that are thus far unconstrained. In addition, the $5\sigma$ discovery reach as a function of $m_\chi$ is a factor of approximately 1.6 in $y$ above the projected exclusion bounds. As a result, DM can be discovered at the $5\sigma$ level with $3000~\ifb$ for DM masses of 3 -- 10 MeV and 50 -- 300 MeV.

\begin{figure}
    \centering
    \includegraphics[width=0.99\textwidth]{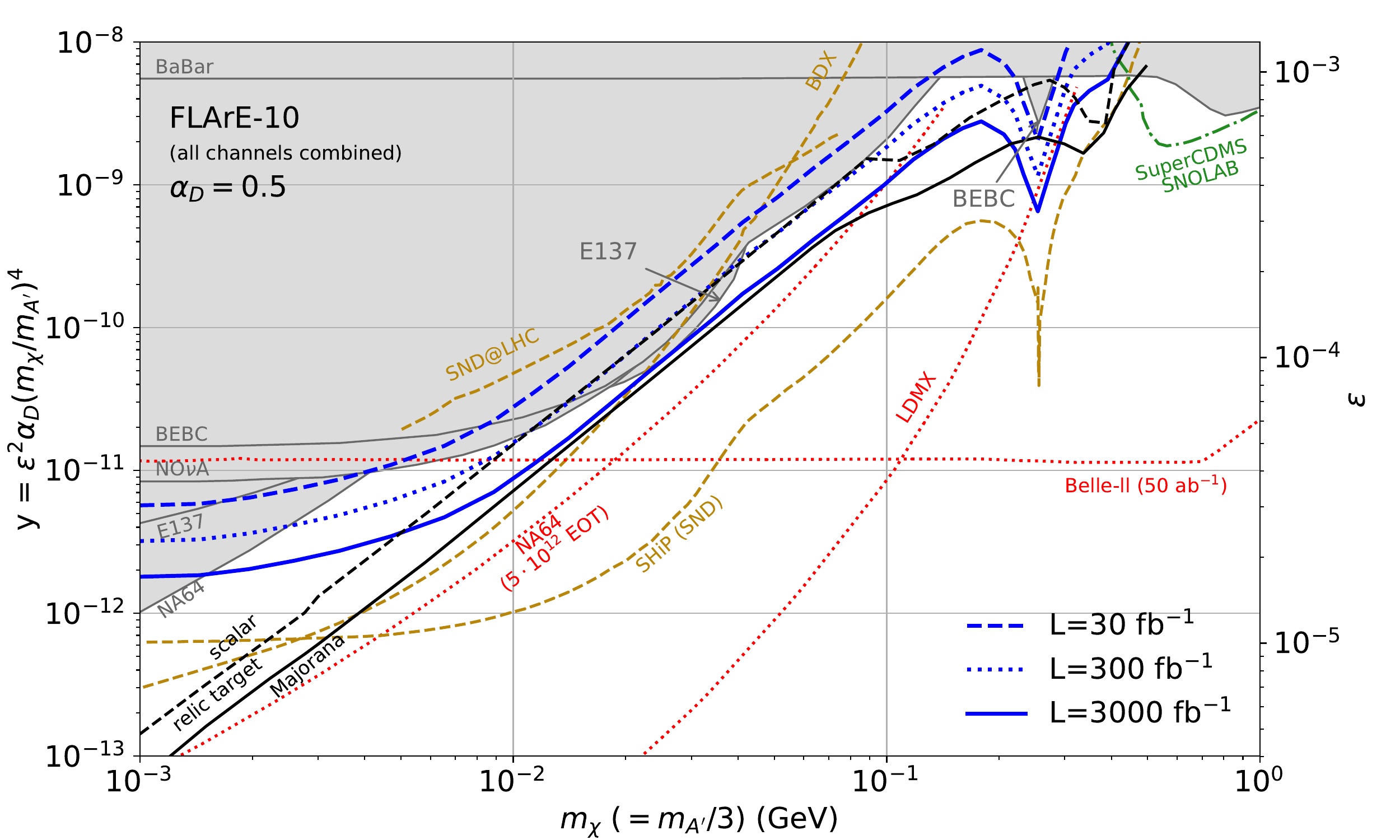}
    \caption{
    The projected $90\%$ CL exclusion bounds for the FLArE-10 detector combining all channels for the three integrated luminosities indicated. New parameter space will start to be probed even for an integrated luminosity of order $30~\ifb$. Existing constraints and projected reaches from other experiments are as in \figref{FL10combined1}.
    }
    \label{fig:combinedFLArE10lumi}
\end{figure}

\section{Conclusions}
\label{sec:conclusions}

The search for terrestrial DM production is a major component of the physics programs of particle accelerator and collider facilities. This avenue is especially useful in the case of sub-GeV DM, where traditional direct detection experiments lose sensitivity. Such light DM at the LHC would be dominantly produced at high rapidities beyond the reach of the general-purpose detectors, motivating dedicated experiments in the far-forward direction. In this work, we have studied potential DM scattering signals in forward detectors at the HL-LHC, as would be possible at the FPF. Our focus has been on interactions between DM and nuclei, complementing previous work on DM-electron scattering.

We have considered detectors based on both emulsion and liquid argon technology. With the use of timing information, it would be possible to reject muon-induced backgrounds, including those from neutral hadron interactions. We thus expect that the dominant backgrounds will be from neutrino scattering. Indeed, the scattering processes that we have considered are analogous to SM processes with neutrinos: elastic scattering, resonant pion production, and DIS. For each of these processes, we have estimated the DM signal and neutrino background, investigating the differences due to kinematics and incorporating the effects of nuclear FSI as appropriate. We find that for DM scattering through a light mediator, it is possible to mitigate neutrino backgrounds with kinematic cuts favoring events with low momentum transfer. This strategy is effective because the heavier weak gauge bosons cause neutrino backgrounds to prefer high $Q^2$ scattering. Similar considerations apply to other signatures, and it would be interesting to study whether additional sensitivity could be obtained with other processes. These include coherent scattering, coherent pion production, and multiple meson production.

Looking at benchmark models with light DM interacting through the minimal dark photon portal, we find new sensitivity in the MeV to GeV mass range. With either complex scalar or Majorana fermion DM, the searches here would test regions of parameter space in which the observed relic density is obtained through thermal freeze-out. As the characteristic energies of the processes that we have studied are different, they have complementary sensitivities. When these searches are combined with those for DM-electron scattering, FASER$\nu$2 and FLArE-10 could cover the relic target for complex scalar DM for DM masses between several MeV and several hundred MeV. FLArE-100 would provide sensitivity to the relic target in a similar mass range for Majorana DM, which is not probed by current experiments. All of these experiments cover much of the parameter space in which the thermal relic density does not overclose the Universe, and they have the potential to provide direct evidence for DM interactions, in contrast to missing momentum-based searches at accelerator and beam dump facilities. Notably, currently unconstrained regions of parameter space can start to be probed with even the first $\mathcal{O}(30)~\ifb$ of integrated luminosity at the HL-LHC. 

The forward region of the LHC offers exciting possibilities to study physics within and beyond the Standard Model. The FPF would extend the reach of the LHC, providing qualitatively new discovery potential in well-motivated theories of light dark sectors. In addition to electron scattering, a suite of nuclear scattering searches at the FPF detectors can be performed to improve our understanding of the nature of DM. In searching for DM and beyond, further exploration of the unique environment provided by the far-forward region at the LHC is warranted to fully leverage collider probes of new physics.

\acknowledgements

We thank Asher Berlin, Milind Diwan, and Anil Thapa for useful discussions.  We are also grateful to the authors and maintainers of many open-source software packages, including BdNMC~\cite{deNiverville:2016rqh}, CRMC~\cite{CRMC}, EPOS-LHC~\cite{Pierog:2013ria}, FeynCalc~\cite{Shtabovenko:2020gxv}, FORESEE~\cite{Kling:2021fwx}, LHAPDF~\cite{Buckley:2014ana}, GENIE~\cite{Andreopoulos:2009rq,Andreopoulos:2015wxa}, and SIBYLL~\cite{Ahn:2009wx, Riehn:2015oba, Riehn:2017mfm}.  The work of B.B.~is supported by the U.S.~Department of Energy under grant No. DE–SC0007914. The work of J.L.F.~is supported in part by U.S.~National Science Foundation Grant No.~PHY-1915005 and by Simons Investigator Award \#376204. A.I.~and R.M.A.~are supported in part by the U.S.~Department of Energy under Grant No.~DE-SC0016013. R.M.A.~is supported in part by the Dr.~Swamy Memorial Scholarship.  F.K.~is supported by the U.S.~Department of Energy under Grant No.~DE-AC02-76SF00515. S.T.~is supported by the grant ``AstroCeNT: Particle Astrophysics Science and Technology Centre'' carried out within the International Research Agendas programme of the Foundation for Polish Science financed by the European Union under the European Regional Development Fund. S.T.~is supported in part by the Polish Ministry of Science and Higher Education through its scholarship for young and outstanding scientists (Decision No.~1190/E-78/STYP/14/2019). S.T.~is also supported in part from the European Union’s Horizon 2020 research and innovation programme under grant agreement No.~962480 (DarkWave project).


\bibliography{references}

\end{document}